\documentclass[12pt,a4paper,english]{article}

\usepackage[bindingoffset=0.3cm,textheight=22.5cm,hdivide={2.7cm,*,2.7cm}, vdivide={*,22cm,*}]{geometry}
\usepackage[bookmarksnumbered=true,breaklinks=true]{hyperref}

\usepackage{amsmath,amsfonts,amssymb,babel,slashed,relsize}

\usepackage[pdftex]{graphicx}

\usepackage[numbers,square,comma,sort&compress]{natbib}

\IfFileExists{dsfont.sty}
	{\usepackage{dsfont}
         \newcommand{\id}{\mathds{1}}}
	{\typeout{Package dsfont.sty was not found, using alternative macros.}
         \let\mathds=\mathbb
         \newcommand{\id}{\mbox{1 \kern-.59em {\rm l}}}}
\let\one=\id

\makeatletter

         \@addtoreset{equation}{section}
\makeatother

\newcommand{\nocontentsline}[3]{}
\newcommand{\tocless}[3]{\bgroup\let\addcontentsline=\nocontentsline#1{#2}#3\egroup}

\newcommand{\qed}{\nobreak \ifvmode \relax \else
      \ifdim\lastskip<1.5em \hskip-\lastskip
      \hskip1.5em plus0em minus0.5em \fi \nobreak
      \vrule height0.75em width0.5em depth0.25em\fi}

\newcommand{\be}{\begin{equation}}
\newcommand{\ee}{\end{equation}}
\newcommand{\eq}[1]{(\ref{#1})}

\newcommand{\bra}[1]{\langle #1|}
\newcommand{\ket}[1]{|#1 \rangle}

\def\nn{\nonumber}
\def\bea{\begin{eqnarray}}
\def\eea{\end{eqnarray}}
\def\obar{\overline}

\def\beqa{\begin{eqnarray}} 
\def\eeqa{\end{eqnarray}} 
\def\beq{\begin{equation}} 
\def\eeq{\end{equation}} 

\def\a{\alpha}          
\def\b{\beta}           
\def\d{\delta}    
\def\e{\epsilon}                
        
\def\g{\gamma}

\def\l{\lambda} \def\L{\Lambda}

\def\s{\sigma}  

\def\th{\theta}

\def\cA{{\cal A}}  \def\cC{{\cal C}}
\def\cD{{\cal D}}  \def\cF{{\cal F}}
 \def\cH{{\cal H}} 
 \def\cK{{\cal K}} 
\def\cM{{\cal M}} \def\cN{{\cal N}} \def\cO{{\cal O}}
 \def\cQ{{\cal Q}} 
  \def\cU{{\cal U}}

\def\mg{\mathfrak{g}}

\newcommand{\R}{\mathds{R}}
\newcommand{\C}{\mathds{C}}

\newcommand{\msu}{\mathfrak{s}\mathfrak{u}}
\newcommand{\mso}{\mathfrak{s}\mathfrak{o}}
\newcommand{\mmu}{\mathfrak{u}}

\def\bit{\begin{itemize}}
\def\eit{\end{itemize}}

\def\({\left(}
\def\){\right)}
\def\diag{\mbox{diag}}

\def\Mat{{\rm Mat}}

\def\d{\delta}

\def\pa{\partial} \def\del{\partial}

\newcommand{\tr}{\mbox{tr}}

\def\bcomment#1{}

\def\LNC{\Lambda_{\rm NC}}
\def\YM{{\rm YM}}

\newcommand{\co}[2]{[#1,#2]}						
\renewcommand{\a}{\alpha}
\renewcommand{\b}{\beta}
\renewcommand{\d}{\delta}

\renewcommand{\th}{\theta}

\renewcommand{\l}{\lambda}

\newcommand{\G}{\Gamma}

\renewcommand{\L}{\Lambda}

\newcommand{\eps}{\varepsilon}

\newcommand{\cf}{cf.\ }

\DeclareMathOperator{\Tr}{Tr}

 \allowdisplaybreaks[3]

\begin{document}

\renewcommand{\title}[1]{\vspace{10mm}\noindent{\Large{\bf
#1}}\vspace{8mm}} \newcommand{\authors}[1]{\noindent{\large
#1}\vspace{5mm}} \newcommand{\address}[1]{{\itshape #1\vspace{2mm}}}


\begin{flushright}
UWThPh-2014-03 
\end{flushright}

\begin{center}

\title{ \Large An extended standard model and its Higgs geometry  \\[1ex]
 from the matrix model }

\vskip 3mm

 \authors{Harold C. Steinacker{\footnote{harold.steinacker@univie.ac.at}}, 
 Jochen Zahn{\footnote{jochen.zahn@univie.ac.at}}
 }
 
\vskip 3mm

 \address{ 
 {\it Faculty of Physics, University of Vienna\\
Boltzmanngasse 5, A-1090 Vienna, Austria  }  
  }


\vskip 1.4cm

\textbf{Abstract}

\end{center}

We find a simple brane configuration in the IKKT matrix model which resembles the standard model at low energies,
with a second Higgs doublet and right-handed neutrinos.
The electroweak sector is realized geometrically in terms of two minimal fuzzy ellipsoids, 
which can be interpreted in terms of four point-branes in the extra dimensions.
The electroweak Higgs connects these branes and is an indispensable part of the geometry.
Fermionic would-be zero modes arise at the intersections with two larger branes, 
leading precisely to the correct chiral matter fields at low energy, along with  
right-handed neutrinos which can acquire a Majorana mass due to a Higgs singlet.
The larger branes give rise to $SU(3)_c$, extended by $U(1)_B$ and another $U(1)$ 
which are anomalous at low energies and expected to disappear.
At higher energies, mirror fermions and additional fields arise, 
completing the full $\cN=4$ supersymmetry.
The brane configuration is a solution of the model, assuming a suitable effective potential and a non-linear stabilization of  
the singlet Higgs. 
The basic results can be carried over to 
$\cN=4$ $SU(N)$ super-Yang-Mills on ordinary Minkowski space with sufficiently large $N$.

\vskip 1.4cm

\newpage

\tableofcontents

\section{Introduction}\label{sec:background}

The main result of this paper is to establish a background of the IKKT or IIB
model \cite{Ishibashi:1996xs} with low-energy physics close to that of the standard model.
This is part of the programme of using matrix models as  basis for a theory of 
fundamental interactions and matter, which has been pursued for many years 
from various points of view 
\cite{Banks:1996vh,Aoki:1999vr,Aoki:1998vn,Taylor:2001vb,Steinacker:2007dq,Steinacker:2010rh,Kim:2011cr}. 
We focus here on the relation with particle physics,
restricting ourselves to the case of flat 4-dimensional space-time.
Indeed it is well-known that flat Minkowski space arises as ``brane'' solution of the IKKT model, 
realized as noncommutative plane $\R^4_\theta$. It is also known that the fluctuations of the 
(bosonic and fermionic) matrices around a background consisting of $N$ coincident such
$\R^4_\theta$ branes give rise to non-commutative maximally supersymmetric
$\cN=4$ $U(N)$ super-Yang-Mills (SYM) on $\R^4_\theta$, cf. \cite{Aoki:1999vr,Steinacker:2010rh}.
Accordingly, our results can be interpreted as statements within noncommutative $\cN=4$ $U(N)$ SYM,
with sufficiently large $N$. In fact, most of the 
results apply also to $\cN=4$ $SU(N)$ super-Yang-Mills on ordinary Minkowski space,
with sufficiently large $N$.
The main difference lies in the $U(1)$ sector, which acquires a special role in 
the matrix model, related to the effective gravity \cite{Steinacker:2007dq,Steinacker:2010rh}; 
however we largely ignore this issue in the present paper. 

At first sight, it may seem hopeless to obtain anything resembling the standard model from 
a maximally supersymmetric gauge theory. However, at low and intermediate energies this can be achieved.
We establish certain backgrounds of the matrix model, interpreted as 
intersecting branes in 6 extra dimensions, which lead to 
fermionic and bosonic low-energy excitations
governed by an effective action which is 
close to the standard model,  with all the correct  quantum numbers.
This is a very remarkable result, given the 
non-chiral nature of $\cN=4$ SYM. 
The price to pay are mirror fermions which arise at higher energies, 
along with Kaluza-Klein towers of massive fields, 
which ultimately complete the full $\cN=4$ spectrum.
There is indeed no way to obtain the standard model without Higgs:
If we switch off the Higgs, some of these mirror modes become (quasi-) massless, 
and combine with the standard model fermions to form
non-chiral multiplets. 
In that respect the Higgs sector differs from the standard model:
It  arises from two doublets which are an intrinsic part of two minimal fuzzy spheres.
The spontaneous symmetry breaking (SSB) pattern is thus more intricate than in the standard model, but this does not 
rule out the possibility that its fluctuations realize the physical Higgs. 
The remarkable point is that the separation into chiral standard-model fields and 
the mirror sector arises quite naturally on simple geometrical backgrounds,
largely reproducing the essential features of the standard model at low energies.

Let us describe the brane configuration in some detail.
Our background consists of
a stack of 3 baryonic branes $\cD_B$ realized as fuzzy spheres (giving rise to $SU(3)_c\times U(1)_B$),
a leptonic brane $\cD_l$, and two other branes $\cD_u$ and $\cD_d$.
These branes are embedded in 6 extra dimensions, such that the standard model fermions arise 
at their intersections. 
The basic mechanism for obtaining chiral fermions on  intersecting non-commutative branes
was already found in \cite{Chatzistavrakidis:2011gs}. 
However in that work, additional intersections led to unwanted fermions with the wrong chiralities, 
and the Higgs was missing. In the present paper, both problems are resolved, 
by realizing the Higgs as intrinsic part of two minimal fuzzy ellipsoids 
(consisting of two quantum cells) which are part of $\cD_u$ and $\cD_d$, respectively.
These ellipsoids intersect $\cD_B$ and $\cD_l$
at their antipodal points, leading to localized chiral fermions.
The electroweak $SU(2)_L$ gauge group arises as the two ``left-handed'' 
intersection loci on $\cD_u$ resp. $\cD_d$ coincide.
This $SU(2)_L$ is broken by the Higgs, which is an intrinsic part of the branes.
This provides a geometrical realization\footnote{The realization in terms of minimal fuzzy ellipsoids 
is in fact somewhat reminiscent of Connes' interpretation of the Higgs connecting two ``branes'' \cite{Connes:1990qp}.} 
of the electroweak symmetry breaking,
which should also protect the Higgs mass to some extent from quantum corrections.
An extra singlet Higgs $S$ connecting $\cD_u$ with $\cD_l$
prevents a right-handed $SU(2)_R$, and breaks lepton number $U(1)_l$.
It should also induce a Majorana mass term for $\nu_R$.

At low energy,
all the 4-dimensional fermions arising on our background 
are  massive Dirac fermions such as electrons\footnote{The neutrinos also arise with a 
right-handed partner.} and massive quarks. Their left- and right-handed chiral components 
transform in different representations of the spontaneously broken (!) gauge group,
coupling to the appropriate gauge bosons.
For example, $e_L$ and $e_R$ arise on two different intersections of the branes, connected by the Higgs.
The Higgs is moreover essential for the  chiral nature of the fermions at the intersections.

 We stress that our results and predictions for the fermionic would-be zero modes 
 arising at the brane intersections are not only theoretical expectations, but
 can be verified numerically. In particular, we can compute the mass spectrum 
given by the eigenvalues of the internal Dirac operator $\slashed{D}_{\rm int}$ on our background,
as well as the approximate localization and chirality of the corresponding fermionic modes. 
The results are consistent with the expectations.
In particular, we clearly see near-zero modes which are localized as predicted on the 
intersecting branes, with the expected chiralities. Their eigenvalues approach 
zero for increasing $N$, with a clear gap to the next eigenvalues corresponding to mirror fermions.  
For a range of parameters we even find good quantitative agreement with our estimates for 
the Yukawa couplings, including the first series of mirror fermions. 

Our brane configuration is a solution of the bare matrix model action, supplemented by  
a simple $SO(6)$-invariant term in the potential. 
Although we add such a term by hand here (thus explicitly breaking supersymmetry), 
it seems plausible that (a more complicated form of) such a potential arises 
in the quantum effective action of the original model.
This reflects the interaction of the branes extended in the extra dimensions,  
due to the conjectured -- and to some extent verified 
\cite{Ishibashi:1996xs,Chepelev:1997av,Alishahiha:1999ci,Taylor:2001vb,Kitazawa:2005ih,Blaschke:2011qu} --
relation with supergravity. 
The singlet Higgs $S$ corresponds to an instability of the linearized wave operator, which we assume to be non-linearly stabilized.

Other ways to obtain chiral fermions in the IKKT model and similar models have been proposed in the literature.
This includes warped extra dimensions \cite{Nishimura:2013moa}, 
allowing to circumvent the index theorem \cite{Steinacker:2013eya}
which applies to product spaces $\R^4_\theta \times \cK$. 
However, no such warped solution of matrix models is known at present.
Chiral fermions can be obtained in unitary matrix models \cite{Aoki:2010gv}, which in a sense have
a built-in toroidal compactification. However these models are not supersymmetric, which may lead to problems 
 upon quantization. In string theory, there are many ways to obtain chiral fermions, however this 
entails the vast landscape of string compactifications with its inherent lack of predictivity.
Avoiding this is in fact one of the main motivations for the IKKT model. Nevertheless, many of the present
ideas related to brane constructions of the standard model  originate from string theory, cf. 
\cite{Berkooz:1996km,Antoniadis:2000ena}.
Finally, it seems likely that a somewhat adapted brane configuration can be found in the BFSS model \cite{Banks:1996vh}.

We should also state the potential problems and pitfalls of our 
proposal. At some scale above the electroweak scale, mirror fermions 
come into play, which couple to the standard model gauge bosons, and may decay into standard model 
fermions via extra massive gauge bosons.
In order to be at least near-realistic, there should be a sufficiently large gap between the 
electroweak scale and the scale of the mirror fermions. 
Unfortunately at tree level, it turns out that this gap is not very large.
However we argue that quantum corrections should increase this gap, since 
a tower of massive Kaluza-Klein gauge bosons couples to the mirror fermions (as well as to the ordinary fermions)
but not to the electroweak gauge bosons or the Higgs.
Proton decay is prevented by baryon number conservation, which is violated only 
by a quantum anomaly.

The solution presented here a priori leads to two generations, which arise from two widely separated
intersection regions of the underlying branes, with the same structure and chiralities. 
It seems straightforward to extend them to any even 
number of generations, by introducing multiple branes. To get an odd number of generations is less clear;
one possibility is that the singlet Higgs $S$ 
leads to a deformation of the background and removes one intersection region.

At this point, 
it is perhaps a bit optimistic to hope that the backgrounds proposed here --  with some adjustments  
 -- can be phenomenologically viable. On the other hand it seems at least conceivable, 
and the fact that we can get so close in this maximally (super)symmetric matrix model is certainly 
very remarkable. 
This should provide motivation to investigate these observations in more detail.

This paper is organized as follows. In section \ref{sec:matrixmodels-intro}, we collect the required facts about the 
matrix model, and recall the relation with noncommutative $\cN=4$ SYM. 
From that point on the paper may be read and interpreted by anyone familiar with $\cN=4$ SYM. 
In section \ref{sec:SM-branes}, the organization of the fermions and their quantum numbers 
is recalled from \cite{Chatzistavrakidis:2011gs}. The central idea of the Higgs realized as intrinsic part 
of a minimal brane is explained in section \ref{sec:minimal-branes}.
In section \ref{sec:intersecting-brane-solutions} we give the  brane solution \eq{branes-explicit} of the matrix model, 
which is the centerpiece of the paper. It is also
spelled out with all branes in \eq{EW-branes}.
The rest of the paper is devoted to establishing the low-energy physics on this background.
The chiral fermions at the brane intersections are established in section \ref{sec:chiral-flat} in the 
flat limit $N \to \infty$, where they become exactly massless. The case of finite $N$
is then discussed in section \ref{sec:Yukawas} using an ansatz motivated by the previous section, 
which allows to estimate the corresponding Yukawa couplings. These are compared with numerical computations.
The symmetry breaking and the resulting low-energy effective field theory is elaborated in section 5,
which allows to make contact with the standard model. In the appendix, we elaborate the reduction of the fermions 
to 4 dimensions.

\section{The matrix model} 
\label{sec:matrixmodels-intro}

Our starting point is the IKKT or IIB model  \cite{Ishibashi:1996xs}, which is given by the action
\begin{align}
S_{\YM} &= \L_0^4\Tr\(\co{X^A}{X^B}\co{X_A}{X_B}\,  +  \obar\Psi \Gamma_A[X^A,\Psi] \)  . 
\label{S-YM}
\end{align}
The indices $A,B$ run from $0$ to $9$, and
are raised or lowered with  the invariant tensor $\eta_{AB}$ of $SO(9,1)$.
The $X^A$ are Hermitian matrices, i.e. operators acting on a separable Hilbert space $\mathcal{H}$, and
$\Psi$ is a matrix-valued Majorana Weyl spinor of $SO(9,1)$, with Clifford generators $\G_A$. 
We also introduced a scale parameter $\L_0$ with $\dim \L_0 = L^{-1}$.
This model enjoys the fundamental gauge symmetry 
\be
X^A \to U^{-1} X^A U\,,\qquad  \Psi \to U^{-1} \Psi U\,,\qquad    U \in U(\cH)\,
\label{gauge-inv}
\ee
as well as the 10-dimensional Poincar\'e symmetry
\be
\begin{array}{lllll}
X^A \to \L(g)^A_B X^B\,,\quad  &\Psi_\a \to \tilde \pi(g)_\a^\b \Psi_\b\,,\quad  & g \in \widetilde{SO}(9,1), &  \\[1ex]
X^A \to X^A + c^A \one\,,\quad & \quad & c^A \in \R^{10}\,\quad 
\end{array}
\label{poincare-inv}
\ee
and a $\cN=2$ matrix supersymmetry \cite{Ishibashi:1996xs}.
The tilde indicates the corresponding spin group.
Defining the matrix Laplacian as
\begin{align}
  \Box \Phi :=  [X_B,[X^B,\Phi]] ,
\end{align}
the equations of motion of the model take the  form
\be
 \Box X^A \,=\, [X_B,[X^B,X^A]] =  0
\label{eom-IKKT}
\ee
for all $A$, assuming  $\Psi = 0$.

\subsection{Noncommutative branes and gauge theory}

We  focus on matrix configurations (in fact solutions, ultimately) which describe embedded
noncommutative (NC) branes. This means that 
the $X^A$ can be interpreted as quantized embedding functions \cite{Steinacker:2010rh}
\be
X^A \sim x^A: \quad \mathcal{M}^{2n}\hookrightarrow \R^{10} 
\ee 
of a $2n$- dimensional manifold embedded in $\R^{10}$. More precisely,
there should be a quantization map $\cQ: \cC(\cM)  \to  \cA\subset L(\cH)$
which maps  functions on $\cM$ to a noncommutative (matrix) algebra, 
such that commutators can be interpreted as quantized Poisson brackets, and $\cA$ as quantized algebra of 
functions on $\cM$.
In the semi-classical limit indicated by $\sim$,  matrices are identified with functions via $\cQ$,
in particular, $X^A = \cQ(x^A) \sim x^A$,
and commutators are replaced by Poisson brackets. For a more extensive introduction
see e.g. \cite{Steinacker:2010rh}. 
Then the commutators
\begin{align}
\co{X^A}{X^B} \ &\sim \  i \{x^A,x^B\} \ = \ i \th^{\a\b}(x)\del_\a x^A \del_\b x^B\,
\end{align}
encode a quantized Poisson structure on $(\cM^{2n},\theta^{\a\b})$.
This Poisson structure sets a typical scale of noncommutativity $\LNC$.
We will assume that $\th^{\a\b}$ is non-degenerate, 
so that the 
inverse matrix $\th^{-1}_{\a\b}$ defines a symplectic form 
on $\mathcal{M}^{2n}\subset\R^{10}$.

The prototype of such a noncommutative brane solution is given by the 4-dimensional quantum 
plane $\R^4_\theta$, defined by $[\bar X^\mu, \bar X^\nu] = i \theta^{\mu\nu} \one$ 
where $\theta^{\mu\nu}$ has rank 4. It obviously satisfies $\Box \bar X^A =0$.
We can  assume that 
this plane is embedded along $\mu = 0,...,3$, with $\bar X^a = 0$ for $a=4,...,9$.
The (well-known) key observation is  that fluctuations of the matrices around this 
background 
\begin{align}
X^A = \bar X^A + \cA^A
\end{align}
describe  (non-commutative) $\cN=4$ gauge theory on $\R^4_\theta$.
Interpreting the fluctuations $\cA^A$ as 
$\mmu(N)$-valued functions in $\R^4_\theta$, 
the matrix model reduces to (cf. \cite{Aoki:1999vr,Steinacker:2010rh})
\begin{align}
 S_{\rm YM} 
 &= \frac{\L_0^4 }{(2\pi)^2} \int d^4 x\, \sqrt{G} 
 \ \tr_N\Big(-\rho^{-1} (\cF \cF)_G 
 - 2 G^{\mu\nu} D_\mu \cA^a D_\nu \cA_a + \rho [\cA^a,\cA^b][\cA_a,\cA_b] \nn\\
   &\qquad \qquad   + \rho^{1/2} \bar\Psi\big(\tilde\g^{\mu}(i\del_\mu + [\cA_\mu,.]) 
   + \rho^{1/2} \Gamma^a [\cA_a,.]\big)\Psi\Big) \nn\\
&=  \int d^4 x\, \sqrt{G} \ \tr_N\Big(-\frac 1{4g_{\rm YM}^2} (\cF \cF)_G 
- \frac 12 G^{\mu\nu} D_\mu \Phi^a D_\nu \Phi_a + \frac 14 g_{\rm YM}^2 [\Phi^a,\Phi^b][\Phi_a,\Phi_b] \nn\\
&\qquad \qquad   + \bar\psi\tilde\g^{\mu}(i\del_\mu + [\cA_\mu,.])\psi + g_{\rm YM} \bar\psi\Gamma^a [\Phi_a,\psi]\Big) 
\label{S-YM-eff}
\end{align}
where 
\begin{align}
 X^\mu &= \bar X^\mu + \theta^{\mu\nu} \cA_\nu ,  \qquad \mu,\nu = 0,...,3 \nn\\ 
G^{\mu\nu} &= \rho \theta^{\mu\nu'}\theta^{\nu\nu'} \eta_{\mu'\nu'} , \nn\\
   \rho &= \sqrt{|\theta^{-1}|},  \nn\\
\Phi^a &=  \frac{\L_0^2}{\pi}\, \cA^a, \qquad\qquad a = 4,..., 9 \nn\\
\psi &= \frac{\L_0^2}{2\pi} \rho^{1/4} \Psi , \nn\\ 
\tilde\g^{\mu} &= \rho^{1/2}\, \theta^{\nu\mu} \gamma_\nu  
\label{field-identification}
\end{align}
and  $\cF_{\mu\nu} = \del_\mu \cA_\nu - \del_\nu\cA_\mu + [\cA_\mu,\cA_\nu]$ is the $\mmu(N)$ field strength.
Since $|G| = |\eta| = 1$ in four dimensions, we will drop $\sqrt{G}$ from now on.
All fields take values in $\mmu(N)$. In particular, we can read off the 
$\mmu(N)$ coupling constant,
\begin{align}
 \frac 1{4 g_{\rm YM}^2} = \frac{\L_0^4}{(2\pi)^2} \rho^{-1}  \ .  
\end{align}
Although the action \eq{S-YM-eff} is written in a way that looks like the standard $\cN=4$ SYM, 
it is in fact noncommutative $\cN=4$ SYM on $\R^4_\theta$. In the present paper, we will focus on those
aspects where this distinction becomes (almost) irrelevant, emphasizing that the basic results
also apply to standard $\cN=4$ SYM on commutative $\R^4$.

To describe the internal or ``extra-dimensional'' sector described by the $\Phi^a$ or $X^{4...9}$, 
we need to consider more general branes $\cM^{2n}$.
Being embedded in $\R^{10}$, they are equipped with the induced metric
\begin{align}
g_{\a\b}(x)=\pa_\a x^A \pa_\b x_A\,
\label{eq:def-induced-metric}
\end{align}
which is the pull-back of $\eta_{AB}$.  However, this is not the effective metric on $\cM^{2n}$. 
It turns out that the  effective action for fields and matter on such NC branes
is governed by a universal effective metric $G^{\a\b}$ given by  \cite{Steinacker:2010rh}
\begin{align}
 G^{\a\b}&= \rho\th^{\a\a'}\th^{\b\b'}g_{\a'\b'}\,,   \qquad 
  \rho  = \Big(\frac{\det{\th^{-1}_{\a\b}}}{\det{g_{\a\b}}}\Big)^{\frac 1{2(n-1)}}   
\label{eff-metric}
\end{align}
for $n>1$. This can be seen  
using the semi-classical form of the matrix Laplace operator\footnote{This result 
does not apply to the 2-dimensional case, where a modified formula holds \cite{Hoppe:2012}.}
\cite{Steinacker:2010rh}
\begin{align}
 {\bf\Box} \Phi &= [X_A,[X^A,\Phi]]  \ \sim\  -\rho^{-1} \Box_{G}\, \phi 
\label{matrix-laplacian-metric}
\end{align}
acting on scalar fields $\Phi \sim \phi$.
Then the matrix equations of motion \eq{eom-IKKT} take the  simple form
\be
0 =  \Box X^A \,\sim\, -\rho^{-1} \Box_{G} x^A ,
\label{eom-IKKT-semi}
\ee
hence the embedding $x^A \sim X^A$ is given by harmonic functions on $\cM$ with respect to $G_{\a\b}$.

The prime example of a compact noncommutative brane is the fuzzy sphere $S^2_N$ \cite{hoppe,Madore:1991bw}. Its
embedding in $\R^3$ is given in terms of 3 matrices
$Y^i = c L^i$, where $L^i$ is the generator of the $N$-dimensional irreducible representation of $\msu(2)$. Then
\begin{align}
  \Box_Y Y^i &= 2 c^2 Y^i , \nn\\
 \sum_{i=1}^3 (Y^i)^2 &= c^2 \frac{N^2-1}{4}
\end{align}
In this paper, we will give such a compactification in terms of stacks of suitable $\cK$,
resulting in a 4-dimensional
gauge theory on $\R^4$ that resembles the standard model at low energy.

The constructions of this paper also apply to $SU(N)$ $\cN=4$ SYM theory
on ordinary $\R^4$. Then the brane configurations become  backgrounds of 
the 6 scalar fields, and our results  state that the low-energy physics of such a background 
resembles that of the standard model.

\section{The standard model from branes in the matrix model}
\label{sec:SM-branes}

\subsection{Fields and symmetries}

In order to recover the standard model from the matrix model, 
all fields must be realized as matrices in the adjoint of some big $U(N)$ gauge group.
The $SU(3)_c \times SU(2)_L \times U(1)_Y$ gauge group must arise
at low energies from the 
fundamental $SU(N)$ gauge group by some symmetry breaking mechanism, and the standard
model matter fields must transform in the appropriate way. 
It is quite remarkable that this is possible at all. Such 
an embedding of the standard model fields was given in \cite{Chatzistavrakidis:2011gs} (cf. \cite{Grosse:2010zq}), 
which we take as starting point here.
The fermionic matrices (including a right-handed neutrino) are realized as follows
\begin{align}
 \Psi=\begin{pmatrix}
  0_{2} & 0 & { \begin{array}{cc} 0 &  l_L \end{array}} & Q_L \\
   &  0  & \begin{pmatrix}
            0 &  e_R \\  0 &  \nu_R 
           \end{pmatrix}
 & Q_R \\
 & & ~~~~0 & 0 \\
&  &  & ~0_{3}
\end{pmatrix},
\label{fermion-matrix}
\end{align}
where
\begin{align}
Q_L = \begin{pmatrix} u_L \\ d_L   \end{pmatrix}, \qquad
l_L = \begin{pmatrix} \nu_L \\ e_L   \end{pmatrix}, \qquad
 Q_R = \begin{pmatrix}
   d_R \\ u_R
\end{pmatrix}
\end{align} 
The electric charge $Q$ and the weak hypercharge $Y$ are then 
realized by the adjoint action of the following $SU(N)$ generators
\begin{align}
t_Q = \frac 12 \begin{pmatrix}
  1 &&&&&\\
  & -1 &&&&\\
  && -1 &&&\\
  &&& 1 && \\
  &&&& 1 &\\
  &&&&& -\frac 13
 \end{pmatrix}, \qquad
t_Y = \begin{pmatrix}
  0 &&&&&\\
  & 0 &&&&\\
  &&-1 &&&\\
  &&& 1 && \\
  &&&& 1 &\\
  &&&&& -\frac 13
 \end{pmatrix} .
 \label{Q-Y-explicit}
\end{align}
In particular, the Gell-Mann Nishima relation 
\begin{align}
 t_Q = t_3 + \frac 12\, t_Y, \qquad
  t_3 = \frac 12\begin{pmatrix}
   1 &&&&&\\
  & - 1  &&&&\\
  && 0 &&&\\
  &&& 0 && \\
  &&&& 0 &\\
  &&&&& 0
 \end{pmatrix}
\end{align}
is satisfied.  
Furthermore, we need a mechanism which breaks the $U(N)$ gauge group down to the standard model 
 gauge group $SU(3)_c \times SU(2)_L \times U(1)$ 
(possibly extended by additional $U(1)$ factors), such that $Q$ and $Y$ arise as above.
This can be achieved naturally by a suitable arrangement of stacks of 
compact branes in the extra dimensions, analogous to brane constructions in string theory \cite{Antoniadis:2000ena}. 
In the matrix model, such a collection of coincident branes $\cK_i$ can be realized by  block matrix configurations 
$X^a_{(i)}$ acting on $\cH_{(i)} \cong \C^{N_i}$, cf. \cite{Aschieri:2006uw}.
This suggests a brane configuration \cite{Chatzistavrakidis:2011gs} with $2+1+1$ ``electroweak'' branes 
$2\times \cD_w\oplus \cD_a \oplus \cD_b$ leading to $SU(2)_L \times U(1)^3$, along with
a ``leptonic'' brane  $\cD_l$ which carries a $U(1)_l$ gauge group
(corresponding to lepton number),
and three coincident ``baryonic'' branes $\cD_B$ which carry the $SU(3)_c \times U(1)_B$  gauge group:
\begin{align}
X^a_{\rm (naive)} =  \begin{pmatrix}
  X^a_{(w)}\otimes\one_2 &&&&\\
  & X^a_{(a)} &&&\\
  &&X^a_{(b)}&&\\
  &&&X^a_{(l)}&\\
  &&&&X^a_{(B)}\otimes\one_3
 \end{pmatrix} .
\end{align}
Here $k$ coincident branes are described by $(...)\otimes \one_{k}$.
This background breaks\footnote{This is nothing but a variant of the usual 
Higgs mechanism, viewing the $X^a$ as scalar fields. We assume that each $X^a_{(i)}$ generates 
the irreducible algebra ${\rm Mat}(N_i,\C)$ of functions on one brane $\cD_i$.} 
the $U(N)$ gauge symmetry down to the product of $U(k_i)$ as follows
\begin{align}
U(N) \rightsquigarrow \diag(U(2)_L,U(1),U(1),U(1),U(3)) .
\label{brane-breaking-naive}
\end{align}
A priori,  fermions  on such noncommutative branes 
are not  chiral, and thus cannot  realize the 
standard model. Remarkably, chiral fermions do arise on intersections of such 
(non-commutative!) branes as shown in \cite{Chatzistavrakidis:2011gs}, 
provided they locally span the  internal space $\R^6$.
Thus for suitable arrangements of the above branes, 
the required chiral fermions \eq{fermion-matrix} may indeed arise on the corresponding intersections of
$\cD_l$ and $\cD_B$ with the electroweak branes $\cD_{w},\cD_a,\cD_b$.
However due to the trivial topology of $\R^{10}$, there are
always additional intersections, leading to additional fermions with the 
opposite chiralities. This is quite unavoidable for branes with
product geometry\footnote{Another possible solution to this problem was proposed in \cite{Nishimura:2013moa}
based on a ``warped'' geometry. However, it is not clear how such configurations can arise in matrix models.} 
$\R^4 \times \cK\subset \R^{10}$,
as can be shown by an index theorem \cite{Steinacker:2013eya}.

We propose a simple solution to this problem here, which at the same time provides a compelling 
mechanism for the electroweak Higgs. 
First, we note  that the Higgs doublets 
\begin{align}
 H_d = \begin{pmatrix}
         0 \\ \phi_d
        \end{pmatrix}, \qquad 
 H_u = \begin{pmatrix}
         \phi_u \\ 0
        \end{pmatrix}
\end{align}
with  $Y(H_d) = 1$ (as in the standard model) and
$Y(H_u) = -1$ (as in the MSSM) fit into the above matrix structure as 
\begin{align}
 X^a_{(H)} = \begin{pmatrix}
         0_2 & H_d & H_u & 0 & 0 \\
         & 0 & 0 & 0 & 0\\
         && 0 & S & 0\\
         &&& 0 & 0 \\
         &&&&0
        \end{pmatrix}
 \ =\ \begin{pmatrix}
          0 & 0 & 0 & \phi_u & 0&0\\
          0&0 & \phi_d & 0 &  0&0\\
           0 & \phi_d^{\dagger} & 0& 0 &0&0 \\
          {\phi_u}^{\dagger} & 0 & 0& 0  & S & 0\\
          0&0&0&  S^\dagger & 0 & 0 \\
          0&0&0&0 & 0 & 0\end{pmatrix} .
\label{higgs-matrix}
\end{align}
This  indeed leads to the desired pattern of electroweak symmetry breaking.
We  also added a ``sterile'' Higgs $S$, which is a  singlet under the 
standard model gauge group, occupying the same slot as $\nu_R$.
This leads to a modified matrix background 
of the form
\begin{align}
 X^a &= X^a_{\rm (naive)} + X^a_{(H)} \ ,
\label{Higgs-matrix-5branes}
\end{align}
which however still 
does not resolve the problem of chirality doubling.
The solution comes from replacing the two branes connected by the Higgs
with a single noncommutative brane, recognizing the Higgs as intrinsic part of the geometry.
This is explained in the next section.

\subsection{Higgs from deconstructing compact branes}

For two branes  connected by an off-diagonal Higgs  as above, the embedding matrices
generate an irreducible algebra which contains the original branes as sub-algebras, 
and should therefore  be interpreted geometrically as a single compact space $\cK$. 
Conversely, a single compact brane $\cK$ 
can be considered as a 2-brane system glued together at the boundary by some Higgs. For example, 
$S^2_N$ can be interpreted as 2 disks in the $xy$ direction near the north and south poles
connected by an equatorial strip, which realizes the  Higgs.
In mathematical terms, we split the Hilbert space of a fuzzy sphere 
$\cH_{N} = \C^N = \C^{N/2} \oplus \C^{N/2} \cong \cH_{(1)} \oplus \cH_{(2)}$
into two halves interpreted as $\cD_{(1)}$ and $\cD_{(2)}$, and write the embedding matrices as
\begin{align}
 X^a = \begin{pmatrix}
        X^a_{(1)} & 0 \\
         0 & X^a_{(2)}
       \end{pmatrix}
 + \begin{pmatrix}
  0 & \phi \\
  \phi^\dagger & 0
   \end{pmatrix} \ .
\end{align}
We can then interpret the two diagonal blocks as 2 a priori separate branes, linked by the Higgs field $\phi$.
Note that $\cD_{(1)}$ and $\cD_{(2)}$ have opposite Poisson structure near the origin, and are transversally 
separated by the diameter. 
\begin{figure}
   \hspace{0.35\textwidth}
\includegraphics[width=0.25\textwidth]{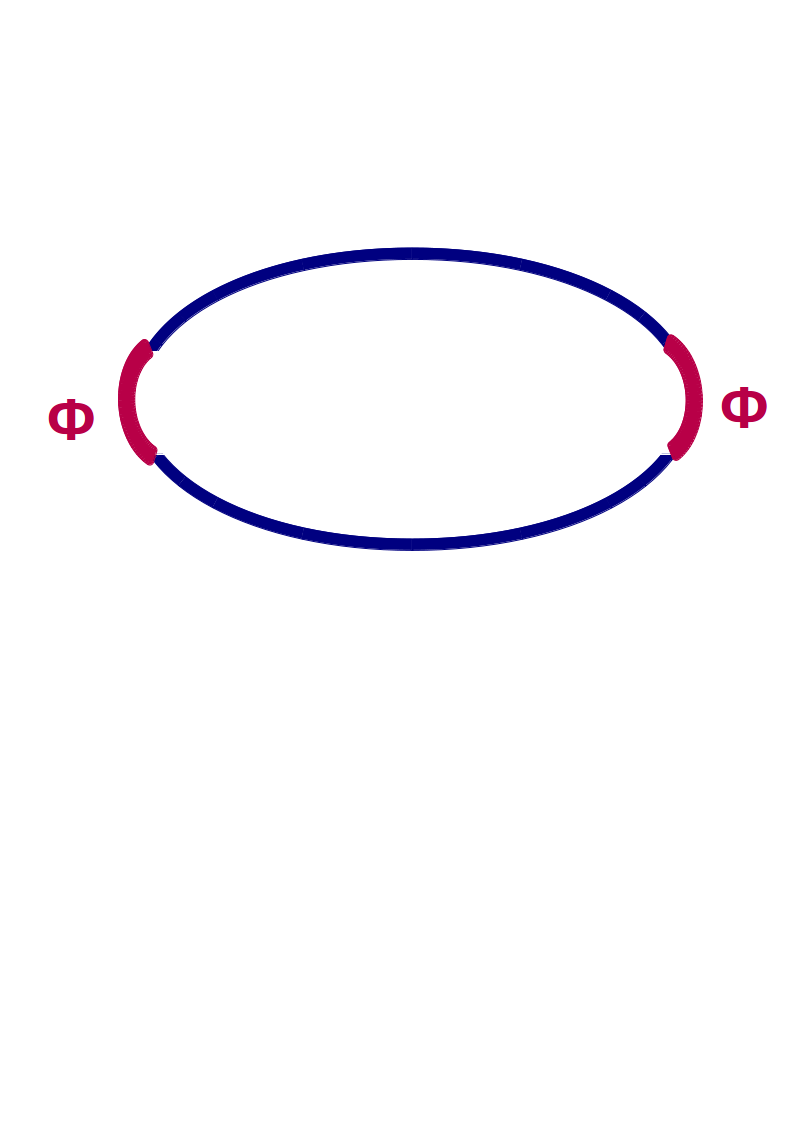}
 \vspace{-0.1\textwidth}
\caption{Higgs from compactified branes}
 \end{figure}
The two  groups $\cU_{(1)} = U(\cH_{(1)})$ and $\cU_{(2)} = U(\cH_{(2)})$ corresponding to the  
diagonal blocks can be viewed as  gauge groups on the two 
half-branes\footnote{They should not be viewed as a stack of identical branes, because
they have opposite orientation.}.
Then $\phi$ intertwines these gauge groups,
and plays the role of a Higgs. Indeed, the 4-dimensional gauge fields corresponding to  
$\cU_{(1)},\cU_{(2)}$ will acquire a mass due the Higgs effect.

One problem with this idea is that the Kaluza-Klein gauge modes on $\cK$
would not respect in general these upper or lower 
half-branes, but  spread over the entire
compact quantum space. Moreover, they would not respect the localized 
fermions arising on intersections of branes in a clear-cut way.
These problems are avoided for fuzzy spaces with $N=2$ represented on $\C^2 = \cH_{(1)} \oplus \cH_{(2)}$ with
$\dim\cH_{(1)},\cH_{(2)} = 1$; we call them minimal $N=2$ quantum spaces.
This leads to a simple set of gauge modes arising from a short KK tower.

\subsection{Minimal electroweak branes with Higgs}
\label{sec:minimal-branes}

\begin{figure}
 \vspace{-0.06\textwidth}
   \hspace{0.3\textwidth}
 \includegraphics[width=0.45\textwidth]{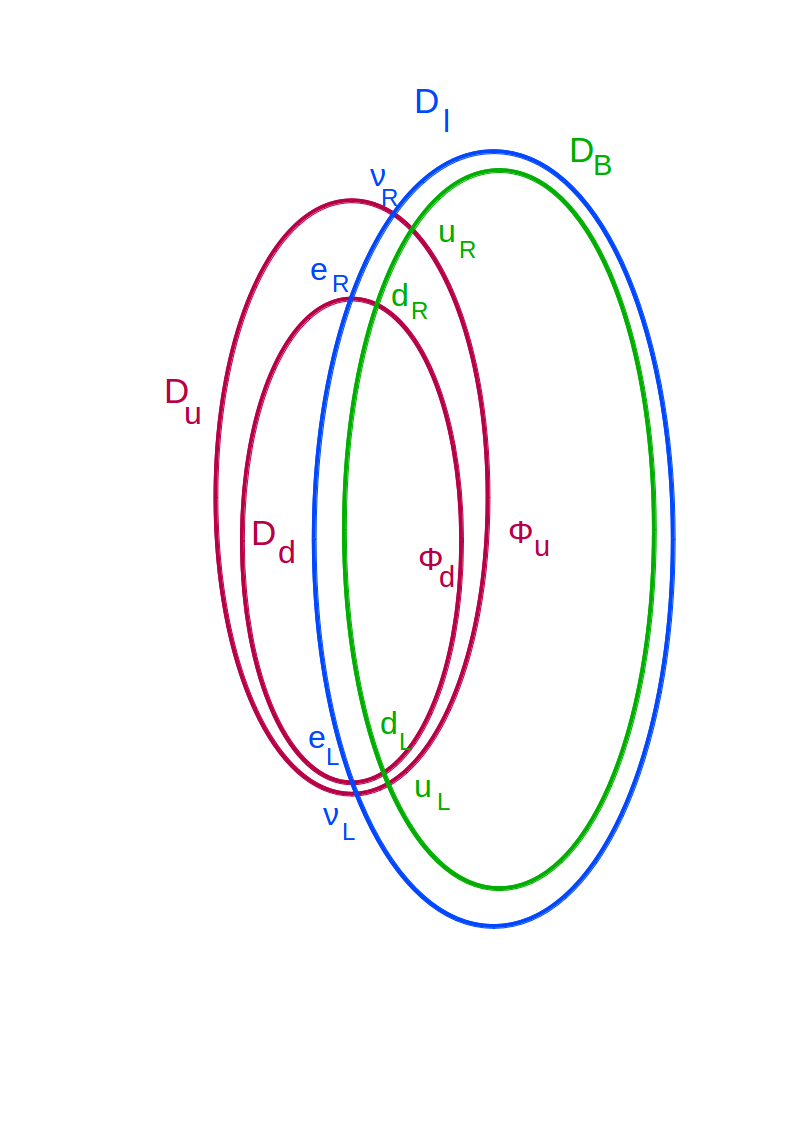}
 \vspace{-0.1\textwidth} 
 \caption{SM brane scheme}
 \label{fig:branes-scheme}
\end{figure}

Applying this idea to the above brane scheme, we interpret the Higgs $\phi_u$  as fusion 
of (the ``half-branes'' defined by) the first and fourth line in \eq{higgs-matrix}
into a single compact brane denoted as $\cD_u$, and $\phi_d$ as fusion of the second and third line
into another compact brane  $\cD_d$.
If these two  branes touch each other at some point, an 
approximate (i.e. spontaneously broken) $U(2)_L$ gauge group 
arises at the intersection, corresponding to
the electroweak $SU(2)_L$ gauge group of the original 
stack of $\cD_w$ branes \eq{brane-breaking-naive}. If that common point of $\cD_u$ and $\cD_d$ is at the intersection with 
$\cD_l$ (and $\cD_B$), then the chiral fermions arising at this location will transform as doublets 
under $SU(2)_L$.  
This leads to a brane scheme as sketched in figure \ref{fig:branes-scheme}.
Although e.g. $\cD_u$  intersects  $\cD_l$ also at another point leading to 
fermions with opposite chirality (as implied by the index theorem), these  fermions now
transform trivially under $SU(2)_L$. 
In this way, an effectively chiral theory can emerge from an underlying non-chiral model.
$SU(2)_L$ is broken by the brane geometry, due to the Higgs identified above as intrinsic part of the brane.

A simple explicit example of such a configuration is given by  two fuzzy spheres embedded 
as follows
\begin{align}
X_4 = \begin{pmatrix}
         \phi_d\l_1 & \\ & \phi_u\l_1
        \end{pmatrix}, \quad 
X_5 =       \begin{pmatrix}
         \phi_d\l_2 & \\ & \phi_u\l_2
        \end{pmatrix}, \quad 
X_6 =   \begin{pmatrix}
         r_d\l_3 & \\ & r_u\l_3 + c \one
        \end{pmatrix}
\end{align}
where $\l_i$ are $\msu(2)$ generators in the $N$-dimensional representation.
For $r_u$, $r_d$ and $c$ appropriately chosen, they touch at the south pole $p_-$, which leads to  
an approximate (spontaneously broken) $SU(2)_L$ at $p_-$
as elaborated
in section  \ref{sec:gauge-fields}.
The corresponding Higgs $\phi$ will be identified shortly. These two fuzzy spheres realize the 
branes $\cD_{u}$ and  $\cD_{d}$ touching each other.
Then $\cD_{u} \cap \cD_B$ intersecting at $p_-$
gives rise to $u_L$ and $\cD_{d} \cap \cD_B$  at $p_-$ gives rise to $d_L$, such that $(u_L, d_L)$
transform as a doublet under $SU(2)_L$. Similarly, 
$\cD_{u} \cap \cD_B$ intersecting at the north pole 
gives $u_R$, and $\cD_{d} \cap \cD_B$ gives $d_R$. These do not transform under $SU(2)_L$.
In the same way, $\cD_{d} \cap \cD_l$  and $\cD_{u} \cap \cD_l$ intersecting at  $p_-$
gives rise to $e_L$ and  $\nu_L$, while  $\nu_R$ and $e_R$ arise at their north poles.
Again, $(\nu_L,e_L)$ form a doublet under $SU(2)_L$, while $\nu_R$ and $e_R$ transform trivially.

\begin{figure}
 \vspace{-0.06\textwidth}
   \hspace{0.3\textwidth}
 \includegraphics[width=0.45\textwidth]{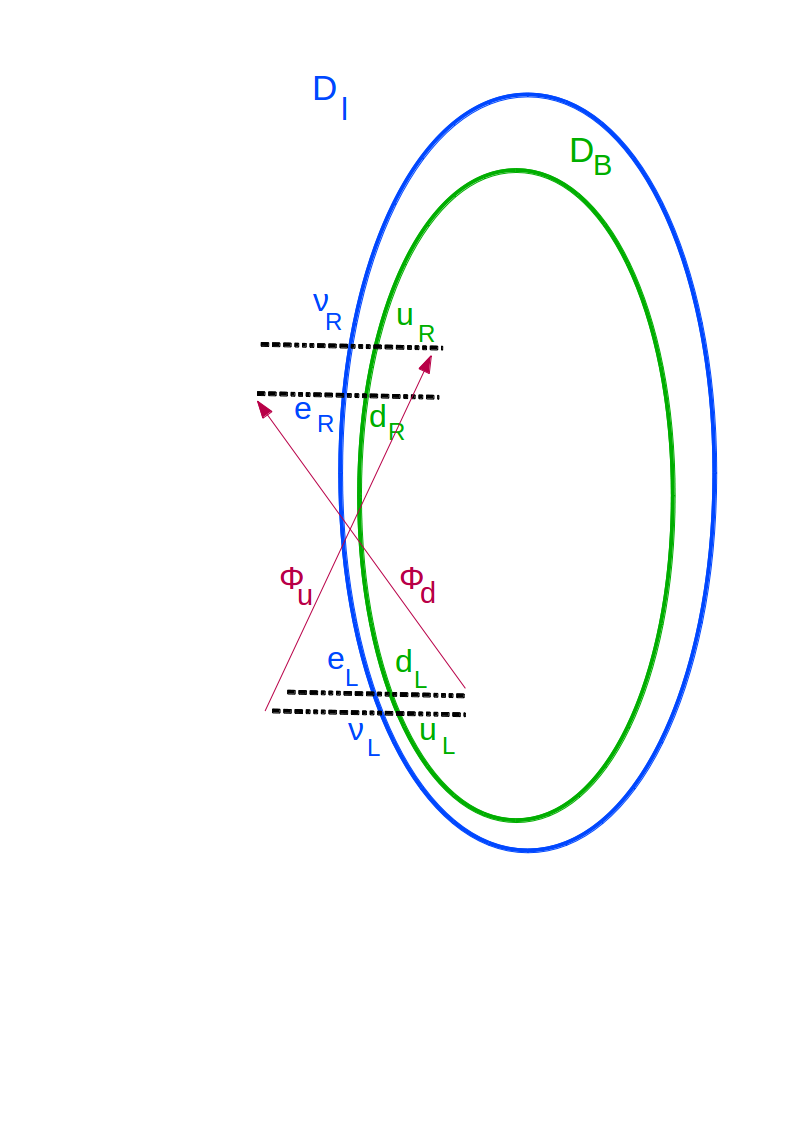}
 \vspace{-0.1\textwidth} 
 \caption{minimal SM brane scheme}
 \label{fig:branes-scheme-N2}
\end{figure}

In general, the physically relevant 4-dimensional gauge fields are determined by a Kaluza-Klein mode expansion on 
$\cD_u$ and $\cD_d$. These modes will in general
not respect the decomposition into upper and lower halves, and couple to fermions with both chiralities
to some extent. 
This problem is resolved if these two fuzzy spheres are realized by $S^2_{N=2}$, 
with minimal Hilbert spaces $\cH \cong \C^2$. 
This leads to the following ``minimal'' electroweak matrix configuration
\begin{align}
X_4 \pm iX_5 &= \frac 12 \begin{pmatrix}
      \phi_d\s^\pm & \\
       & \phi_u\s^\pm
     \end{pmatrix} , \\ 
X_6 & = \frac 12\begin{pmatrix}
       r_d\s_3   & \\
       &  r_u\s_3 + c \one
      \end{pmatrix} 
      = \frac 12\begin{pmatrix}
         r_d & & & \\
         & -r_d & &  \\
         & & r_u + c  & \\
         & & & -r_u + c
        \end{pmatrix}
 \label{branes-scheme-N2}
\end{align}
visualized in figure \ref{fig:branes-scheme-N2}. 
Note that $X^6$ has four eigenvalues,  two of which coincide if $c = r_u - r_d$. 
In the absence of $\phi$, the unbroken gauge group given by the 
commutant (stabilizer) of this background is therefore $U(2)_L\times U(1)\times U(1)$ 
in that case\footnote{The fate of the various $U(1)$ factors will be discussed in section \ref{sec:gauge-fields}.}. 
The gauge modes which do not commute with $X^6$
acquire a  mass $m^2 \sim[X_6,[X_6,.]] $ given by the difference of the $X^6$ eigenvalues.
The $U(2)_L$ is broken in the presence of $\phi$, which will play the role of the 
electroweak Higgs. 
Furthermore, it turns out that chiral fermionic zero modes arise at the intersections 
 even for this very fuzzy geometry, realized by coherent states $|\pm\rangle_{u,d}$ on the branes
 located at the poles $p_\pm$. This will be verified explicitly in section \ref{sec:chiral-flat}.
These fermions couple
to the low-energy gauge group as required, and are connected by the  $\phi_{u,d}$.
Although these $\phi_{u,d}$ clearly correspond to the electroweak Higgs sector, 
the precise role of their fluctuations and the relevance of the other geometrical moduli in the above configuration 
remains to be clarified.

It turns out that in order to have a configuration which is a solution to our modified action, 
we have to take $r_d = r_u$ and $c=0$. Then one also has a broken $U(2)_R$ at the north pole. 
This will be broken not only by the above Higgs $\phi$, but also 
at a higher energy scale by a non-vanishing expectation value of a singlet Higgs $S$. 
It connects $\cD_u$ and $\cD_l$ at the north pole of $\cD_u$, thus lifting the degeneracy of the north poles 
of $\cD_u$ and $\cD_d$. This is elaborated below. In particular,  the breaking of 
the right-handed $SU(2)_R$ is discussed in 
section \ref{sec:gauge-fields}.

\paragraph{Discussion.}

Before establishing these claims in more detail, we briefly discuss some of the further issues
arising in this scenario.

We need to specify the dimensions and type of the various branes.
First, the above remark suggests that all four electroweak D0 branes corresponding to
$|\pm\rangle_{u,d}$ should be located on both branes 
$\cD_B$ and $\cD_l$, in order to obtain chiral near-zero modes which
couple appropriately to the electroweak gauge fields. 
This suggests that $\cD_B$ and $\cD_l$ should be (nearly) coincident.

To get chiral fermions, the branes must span the internal space $\R^6$ at the intersections.
Thus we have two possibilities: either the electroweak branes $\cD_{u,d}$ are two-dimensional 
and $\cD_{l,B}$ are four-dimensional, or conversely. This choice affects the effective 4-dimensional 
gauge couplings, via the volume or trace over the extra dimensions.
It turns out that the first possibility 
leads to a pattern of the electroweak gauge coupling constants (in particular the Weinberg angle) 
which seems unrealistic. 
We therefore take $\cD_{l,B}$ to be 2-dimensional fuzzy branes $K_{N_2}$ with large $N_2$, 
while the $\cD_{u,d}$ have the structure $ K_{N_1} \times S^2_{N=2}$. The
extra $K_{N_1}$ does not significantly change the above picture
of the electroweak symmetry breaking, and merely introduces a multiplicative factor to the 
low-energy gauge groups. This allows a reasonable pattern of low-energy 
coupling constants, as discussed in section \ref{sec:gauge-fields}.

An important question is the fate of the extra $U(1)$ gauge fields,
which always arise in similar brane constructions \cite{Antoniadis:2000ena}.
Each brane comes with an associated
$U(1)$ acting with $\one_i$ on $\cH_{(i)}$, which  do not acquire any mass term from 
a Higgs mechanism. The  trace-$U(1)$ decouples completely in the commutative limit 
(i.e. for ordinary $\cN=4$ SYM), 
and can be identified with a gravitational mode on non-commutative space-time \cite{Steinacker:2010rh,Steinacker:2007dq}; 
we will therefore ignore it in the present paper.
Furthermore a $U(1)_B \sim \frac 13 \one_B$ corresponding to baryon number $B$ arises on the baryonic brane $\cD_B$,
and a $U(1)_l \sim \one_l$ corresponding to lepton number $l$ arises on the leptonic brane $\cD_l$. 
Some of these will be affected by quantum anomalies, as discussed later.
Most importantly, the electric charge 
\begin{align}
 t_Q = \frac 12(\one_u-\one_d + \one_l-\frac 13\one_B)
\end{align}
also arises in this way, which is of course anomaly-free.

Finally, quantum effects are expected to play an important role. 
Besides introducing (benign)
anomalies, they will also mediate the interaction between the branes, which is expected to
play an essential role in selecting and stabilizing the appropriate brane configuration.
This should be a central theme for future work in this context.

\paragraph{Singlet Higgs $S$.}

To avoid an exactly massless $U(1)_{B-l}$ gauge field
and to break $SU(2)_R$ for $r_u = r_d$, we  assume that there is an extra singlet Higgs $S$
connecting $\cD_u$ with $\cD_l$ at the location of $\nu_R$.
$S$ can be seen as superpartner of  $\nu_R$, and it is a singlet  of the standard model gauge group
(cf. \eq{Q-Y-explicit}).
In the presence of a VEV $\langle S\rangle \neq 0$, the $\cD_l$ and  $\cD_u$ branes are
unified into a single compact brane, which is natural in view of 
$t_Q= \frac 12(\one_u+\one_l - ...)$.
Clearly $\langle S\rangle$ breaks $U(1)_{B-l}$, leaving only 
one extra $U(1)_{5}$ gauge field besides the standard model gauge group at low energies. 
That $U(1)_5$  acquires a mass by the electroweak Higgs, and is anomalous at low energies.
Such anomalous $U(1)$ gauge fields are expected to disappear from the low-energy spectrum
by some variant of the St\"uckelberg mechanism, as discussed e.g. in 
\cite{Preskill:1990fr,Coriano':2005js,Morelli:2009ev}.
The symmetry breaking will be discussed in more detail in section \ref{sec:gauge-fields}.

Finally, $\langle S\rangle \neq 0$ allows to write down a Majorana mass term 
for the right-handed neutrino, such as
\begin{align}
 \int d^4 x\,\tr_N(\nu_R^T \g^0 S^\dagger \nu_R S^\dagger) .
 \label{majorana-mass-nu}
\end{align}
Such a term is compatible with the 
full $SU(N)$ gauge symmetry, and could therefore arise in the quantum effective action
even at high scales.

\subsection{Intersecting brane solutions}
\label{sec:intersecting-brane-solutions}

In general, compact quantum spaces in Euclidean signature are never solutions
of the classical matrix equation of motion $\Box X^a = 0$. 
However, quantum effects will contribute to the effective action. 
It is generally expected that this can be related to some sort of 
effective (super-) gravity in higher dimensions; for some partial results from the matrix model point of view 
see e.g. \cite{Ishibashi:1996xs,Chepelev:1997av,Alishahiha:1999ci,Taylor:2001vb,Kitazawa:2005ih,Blaschke:2011qu}. 
In particular, this should lead to an attractive interaction between nearly-coincident branes,
and it is plausible that suitable compact brane configurations may be stabilized in this way.
Lacking more specific results, we will model these quantum contributions to the effective potential
on a 4-dimensional space-time $\R^4_\theta$ 
by a $SO(6)$-invariant function $f\big(\tr_N\sum_{a=4}^9 X_a X^a\big)$:
\begin{align}
\label{eq:S_mod}
S_{\rm YM} &\to S_{\rm YM} - \rho \int d^4 x\, \sqrt{G} \, V_{\rm quant}, \\
\label{eq:V_quant}
V_{\rm quant} &= f(\L_0^2\tr_N\sum_{a=4}^9 X_a X^a) 
 = f(\pi  \rho^{-\frac 12} g_{\rm YM}\tr_N\sum_{a=4}^9 \Phi_a \Phi^a).
\end{align}
Note that the non-commutativity scale $\rho$
allows to write down a dimensionless invariant radius operator.
This leads to the equations of motion\footnote{Note that the regularization for the matrix model proposed in 
\cite{Kim:2011cr} also leads to the same type of equations of motion.}
\begin{align}
 \Box_X X^a = - 2 \pi g_{\YM} \rho^{-\frac 12} f' X^a ,
 \label{eom-quantum}
\end{align}
which will have non-trivial brane solutions (reflecting the above discussion)
provided $f'<0$ in some range.
In particular, we give a solution with the properties discussed above,
where $\cD_l$ and $\cD_B$ are realized as fuzzy spheres 
$S^2_{N_l}$ and a stack of three $S^2_{N_B}$, while the electroweak branes $\cD_u$ and $\cD_d$
are realized as $S^2_{N_u} \times S^2_{N=2}$ and $S^2_{N_d} \times S^2_{N=2}$.

Recall that a fuzzy sphere $S^2_N$ is the matrix algebra $\Mat(N, \C)$ generated by the spin $\frac {N-1}2$ 
representation of $\msu(2)$
\[
 [L_i, L_j] = i \eps_{ijk} L_k \ , 
\]
with radius $\frac 12\sqrt{N^2-1} \sim N/2$. Denote the generators of $S^2_{N_u}$ by $L_i$ and those of 
$S^2_{N_l}$ by $K_i$. The generators of $S^2_2$ are $\sigma'_i$, which are the Pauli matrices $\sigma_i$ divided by 2. 
We also use the notation $L_\pm = L_1 \pm i L_2$. Now let $\cD_u = S^2_{N_u} \times S^2_2$ and $\cD_l = S^2_{N_l}$ be 
embedded as
\begin{align} 
 X^a_{(u)} = \begin{pmatrix}
  R'_u L_3 + \phi_u \s_1' \one_{N_u}  \\
  \phi_u \s_2'\one_{N_u} \\
  r_u \s_3' \one_{N_u} \\
  0 \\ 
  R_u L_1 \\
  R_u L_2 
 \end{pmatrix}, \qquad 
 X^a_{(l)} =  \begin{pmatrix}
      R_l' K_3 \\
     0 \\
      R_l K_1  \\
      R_l K_2  \\
     0 \\
     0
    \end{pmatrix} .
 \label{branes-explicit}
\end{align}
and analogously for $\cD_d$ and $\cD_B$. 
This defines the basic background solutions under consideration here.
The equations of motion  \eq{eom-quantum} are satisfied provided
\begin{align}
 2 R_u^2 = r_u^2+\phi_u^2 = 2 R_l^2
 = R_l^2+ {R_l'}^2 = 2 \phi_u^2 = {R_u'}^2 + R_u^2
 = - 2 \pi g_{\YM} \rho^{-\frac 12} f'
 \label{eom-solution}
\end{align}
which implies 
\begin{align}
  R_u^2 = {R_u'}^2 =  R_l^2 = {R_l'}^2  = r_u^2 = \phi_u^2 = - \pi g_{\YM} \rho^{-\frac 12} f' 
 \label{eom-solution-2}
\end{align}
and similarly for the other branes.
Nevertheless, since the above effective action is certainly oversimplified, we will 
keep the different geometrical moduli $r_i,\phi_i, R_i$ henceforth,  assuming only that they 
have the same scale.

The above brane configuration can alternatively be obtained as solution of the bare
$\cN=4$ SYM equations of motion on $\R^4$ without  quantum corrections, by letting them rotate 
in the extra dimensions as
$X^a = {\bf R}^a_b(t) \bar X^b$, where $\bar X^a$ is given by \eq{branes-explicit}, 
cf.  \cite{Steinacker:2011wb,Hoppe:1997gr}. 
This is indeed a solution for suitable rotations in the $4-5$, $6-7$ and $8-9$
planes.
However the rotation may distort the low-energy effective field theory in 4 dimensions, and we 
will not pursue this possibility here.

\paragraph{Intersections.}

Now consider the intersections of these branes.
If $r,\phi \ll R N_{u,l}$, then  $\cD_u$ and $\cD_l$  intersect near $L_\pm \approx 0 \approx K_\pm$,
provided $R'_u L_3 \approx R'_2 K_3$ up to corrections of order $\phi$. This requires $R'_2 N_l \approx R'_u N_u$.
There are hence two widely separated intersection regions located in target space 
approximately at $\pm  R'_2 N_l (1, 0, 0,0,0,0)$.
Since the spheres are oriented,
the helicity of the would-be zero modes is the same in the two intersection regions, 
as discussed in section \ref{sec:chiral-flat}.
These two intersection regions could therefore be interpreted in terms of two generations.
Alternatively, a deformation of $\cD_l$ (e.g. by the singlet Higgs $S$) might remove one of these intersection regions,
leaving only one generation at this stage\footnote{For example, this is 
achieved by slightly shifting one end of $\cD_l$.}.

For the intersections of the other brane pairs, we need similarly
$R'_2 N_l \approx R'_d N_d \approx R'_2 N_B$.
We will also impose $N_u = N_d$, so that $SU(2)_L$ can act naturally on the Hilbert spaces 
of $\cH_u$ and $\cH_d$, as discussed below. To satisfy all 
these conditions\footnote{As discussed later, one way to introduce additional generations is via
additional branes $\cD_{u,d}^{(i)}$. Their parameters $R^{(i)}$ and
their quantum numbers $N^{(i)}$ should be very close to each other to ensure that they all intersect with the same
$\cD_{l,B}$ branes. This leads to further constraints, and 
different $N^{(i)}$ are possible only if $R'_{u,d} \leq \phi$.}, 
it follows that $N_l \approx N_B$, and $R_u' \approx R_d'$, so that $\cD_u \approx \cD_d$ and $\cD_l \approx \cD_B$
from a geometric point of view.

\subsection{Flat limit $S^2_N \to \R^2_\theta$}

To understand the intersections discussed above for small but finite $r,\phi$ in a simple way, we 
want to approximate the large fuzzy spheres near these intersections by tangential quantum planes. 
We will thus replace $\cD_u$ by $\R^2_{89} \times S^2_2$ 
and $\cD_l$ by $\R^2_{67}$.
In the limit of large $N$, the tangent space to a ``point'' on the fuzzy sphere generated by $R L_i$ tends indeed
to the quantum plane $\R^2_\theta$, if accompanied by a suitable scaling of $R$. As the number of Planck 
cells is $N$ and the area is proportional to $R^2 N^2$, $R$ should scale as $R \sim N^{-\frac{1}{2}}$ in order to have 
a constant Planck cell size and thus a well-defined flat limit. However, in the above configuration, 
a scaling of $R_{u,l}^{(')}$ would have to be accompanied by a scaling of $\phi, r$. Hence, in order to keep $\phi, r$ constant, 
we keep $R_{u,l}^{(')}$ constant, and thus obtain a quantum plane with noncommutativity $\theta \sim N$. 
Specifically, we can replace the tangent spaces of the large spheres by quantum planes 
 \begin{align}
 L_\pm & \to \sqrt{N_u}\, (y^8 \pm i y^9) , 
 & L_3 & \to \pm \frac{N_u}2 \Big(1 - \frac 2{N_u^2} (y_8^2 + y_9^2)  + \cO(\frac yN)^4 \Big), \\
K_\pm & \to \sqrt{N_l}\, (y^6 \pm i y^7) , 
  & K_3 & \to \pm \frac{N_l}2  \Big(1 - \frac 2{N_u^2} (y_6^2 + y_7^2) + \cO(\frac yN)^4 \Big),
 \end{align}
embedded in the $8-9$ and the $6-7$ directions, 
where the $y^i$ fulfill standard commutation relations
\begin{align*}
 [y^6,y^7] & = \pm i, &
 [y^8,y^9] & = \pm i.
\end{align*}
Here the sign depends on the sign of $L_3$ and $K_3$ respectively, and determines the chiralities of 
the would-be zero modes.
Then the effective noncommutativity of the tangential generators $x^a = R \sqrt{N} y^a$
is given by $\theta \sim  R^2 N$, and e.g. the equation of motion \eq{eom-solution-2} for the $8-9$
components becomes
\begin{align}
 \frac 1N \theta_{89} \Big(1+\frac{(R_u')^2}{R_u^2}\Big) = -2 \pi g_{\rm YM} f'\rho^{-\frac 12} .
 \label{omega89-eom}
\end{align}
We can now describe the intersections of $\cD_u$ with $\cD_l$ in more detail, assuming $r,\phi \ll N R$.
In the limit of large $N$, we can write
\begin{align}
 X^a_{(u)} = \begin{pmatrix}
  \pm \frac 12 R'_u N_u + \phi \s_1'  \\
  \phi \s_2' \\
  r \s_3' \\
  0 \\ 
  R_u\sqrt{N_u}\, y^8 \\
  R_u\sqrt{N_u}\, y^9
 \end{pmatrix}, \qquad 
 X^a_{(l)} =  \begin{pmatrix}
     \pm \frac 12 R_l' N_l \\
     0 \\
      R_l \sqrt{N_l}\, y^6 \\
      R_l \sqrt{N_l}\, y^7 \\
     0 \\
     0
    \end{pmatrix} .
\end{align}
Assuming $R'_u N_u = R_2' N_l$ to have perpendicular intersections, 
this reduces to the intersections of a minimal ellipsoid with a quantum plane, 
$\R^2_{\theta (89)} \times S^2_2 \cap \R^2_{\theta (67)}$. The picture of intersecting branes 
makes sense even for minimal fuzzy spheres $S^2_{N=2}$, since their coherent states are located at 
the corresponding classical ellipsoid 
\begin{align}
 \frac{x_4^2 + x_5^2}{\phi^2} +  \frac{x_6^2}{r^2} = 1 .
\end{align}
Taking into account the curvature of $\cD_l$ near $y=0$,
the intersection is determined by the $456$ coordinates
\begin{align}
\begin{pmatrix}
   N_u R_u' + \phi \sin\varphi \\
  0 \\
  r \cos\varphi
\end{pmatrix} 
=  \begin{pmatrix}
    N_l R_l' \cos\vartheta  \\
   0 \\
   N_l R_l \sin\vartheta
  \end{pmatrix}
 \label{intersect-class}
\end{align}
where $\varphi$ is the angle on the normalized minimal fuzzy sphere,
and $\vartheta$ on the large circle of $\cD_l$. 
This suggests the following ansatz for the would-be zero modes
\begin{align}
 \psi = |\varphi +\rangle_u\langle \vartheta|_l  \otimes |s_{\underline{i}}\rangle
 \label{zeromodes-spheres-ansatz}
\end{align}
in terms of coherent states  located at their classical intersection; 
here $|\varphi+\rangle_u$ is the product of coherent states\footnote{Coherent states on fuzzy spheres
are obtained by $SO(3)$ rotations of the highest weight states, cf. \cite{Perelomov:1986tf}.} 
located at the angle $\varphi$ on $S^2_2$ and at the north pole
$L_3 = +\frac{N_u-1}2$ of $S^2_{N_u}$,
$\langle\vartheta|_{l}$ is a coherent state on $S^2_{N_l}$ located at the angle $\vartheta$, and
$|s_{\underline{i}}\rangle$ indicates a suitable spinor state.
It is not hard to see that this leads to approximate zero modes, consistent with the 
picture expected from the flat limit. 
However, we largely restrict ourselves to the flat limit in this paper, as elaborated below.

\subsection{The singlet Higgs $S$.}
\label{sec:singlet-Higgs}

To complete the background,
we have to discuss the
singlet Higgs $S$, linking $\cD_u$ with $\cD_l$ at  $\nu_R$.
Such a link between two branes will be localized at one (or both) of their intersections, suggesting an ansatz 
\begin{align}
H^a_{(S)} &= h^a S \ + h.c., \qquad 
S = \sum_n \ket{p_n +}_u \bra{q_n}_{l}  .
\label{S-ansatz}
\end{align}
Here $\ket{p+}_u$ denotes the tensor product of a state $p$ on $S^2_2$ and the coherent state 
$L_3 \ket{+} = \frac{N_u-1}{2} \ket{+}$ on $S^2_{N_u}$, and $q$ is a state on $S^2_{N_l}$.
For $h$, we choose the ansatz
\begin{align}
 h^a = h (e^8 + i e^9),
 \label{S-polarization}
\end{align}
so that
\begin{equation}
\label{eq:Xh}
 X^a h_a =  h R_u L_+.
\end{equation}
Now we study the linearized wave operator on the perturbation $H^a$, which reads
\begin{equation}
\label{eom-Higgs}
 (P H)^a =  [X^b, [X_b, H^a]] + 2 [[X^a, X^b], H_b] - [X^a, [X^b, H_b]] + 2\pi g_{\rm YM} \rho^{-\frac 12} f' H^a.
\end{equation}
Due to \eqref{eq:Xh}, we have
\begin{equation*}
 [X^b, h_b S] =  h R_u e^{i \omega_S t} L_+ \sum_n \ket{p_n +}_u \bra{q_n}_l = 0.
\end{equation*}
Similarly, we compute for $a \in \{ 4, 5,6,7 \}$
\[
 [[X^a, X^b], h_b S] =  R_u R'_u h  e^{i\omega_S t} \delta^a_4 L_+ \sum_n \ket{p_n +}_u \bra{q_n}_l = 0.
\]
It follows that for $a \not\in \{ 8, 9 \}$, the equation of motion \eqref{eom-Higgs} is fulfilled. 
For $a \in \{ 8, 9 \}$, we note that
\begin{align*}
 [[X^a, X^b], h_b S] & = - \frac 12 R_u^2 (N_u-1) h^a S, \\
 \sum_{b=8}^9 [X^b,[X^b,h^a S]] & = \frac 12 R_u^2 (N_u - 1) h^a S.
\end{align*}
We choose the state $\sum_n p_n \otimes q_n$ such that
\begin{align}
  \sum_{b=4}^7 [X^b,[X^b, S]] = \lambda  S \ .
  \label{eom-lambda-S}
\end{align}
The double commutator on the lhs is a hermitian operator on $\C^2 \otimes \C^{N_l}$, which can be diagonalized. 
One would expect that the two lowest eigenstates $\l$ are localized near the intersections 
$(0,0,\pm 1)$ on $S^2_2$ and $S^2_{N_l}$, suggesting the ansatz $S = \ket{\varphi +}_u \bra{\vartheta}_l$. Then one expects
\begin{align}
\l \ \approx \  \frac 12\phi^2 + \frac 12 R_l^2 (N_l-1) 
\label{omega-S}
\end{align}
for large $N$. This result, and also the localization, can be checked numerically with very high accuracy.
Choosing one of these corresponds to either 
coupling the right- or the left-handed neutrino to the scalar Higgs.
For the action of the wave operator  \eq{eom-Higgs} on \eq{S-ansatz}, we thus obtain
\begin{align*}
 (P H)^a & = \left( -  \frac{1}{2} R_u^2 (N_u-1) + 2\pi\rho^{-\frac 12} g_{\rm YM} f' + \lambda \right) H^a \\
 & \approx \left( \frac{1}{2} R_u^2 (N_l - N_u) - \frac{3}{2} \phi^2 \right) H^a,
\end{align*}
where we used \eqref{eom-solution-2}. In order to have the correct intersection, we need $N_l = N_u$. 
Hence,  our ansatz \eqref{S-ansatz} corresponds to a negative mode of the linearized wave operator,
i.e., an instability. In the following, we assume that it is nonlinearly stabilized, 
so that $h$ acquires a nontrivial value. We plan to address this issue in a forthcoming paper.


The seemingly ad-hoc coupling of $S$ to $\cD_u$ rather than $\cD_d$ can be interpreted as 
spontaneous breaking of the $SU(2)_L \times SU(2)_R$ gauge symmetry down to $SU(2)_L\times U(1)$.
This is discussed further in section \ref{sec:gauge-fields}.
Furthermore, the back-reaction of $S$ to $\cD_{u,l}$ 
might lead to a shift of the branes, possibly removing the other intersection regions
of the  branes (e.g. at $X^4 \approx -RN$). 
Then a single generation would arise for the above background.



\section{Chiral fermions in the flat limit $N \to\infty$}
\label{sec:chiral-flat}

In this section, we consider the limit $N \to \infty$, where the fuzzy spheres $S^2_N$ become 
much larger than the minimal electroweak branes $S^2_{N=2}$. 
We can then replace $S^2_N$ by a quantum plane $\R^2_\theta$ near the intersection with $S^2_2$, and 
obtain exact  results for the (would-be) chiral fermions.
This allows to understand the resulting low energy physics in a simple way.

\subsection{$S^2$ intersecting $\R^2$}
\label{sec:intersecting-plane}

We  want to understand the origin of massless chiral fermions arising on the intersection of the
above minimal ellipsoids embedded in the $456$ directions
with a flat brane $\R^2_\theta$ in the $67$ plane, dropping the $89$ directions for now.
Thus  consider $\cD_u$ realized by  $Mat(2, \C)$ acting on $\cH_{(u)} \cong \C^2$, and
$\cD_l$ realized by an operator algebra acting on  $\cH_{(l)}$.
We should therefore find the (near-) zero modes of the ``internal'' Dirac operator 
in the $4567$ direction 
\begin{align}
  \slashed{D}_{\rm int} \Psi &= \sum_{a=4}^7 \Delta_a [X_a,\Psi] 
   =  \sum_{a=4}^7 \Delta_a (X^a_{(u)} \Psi -  \Psi X^a_{(l)})  
\end{align}
(using the conventions of appendix \ref{sec:clifford})
for the off-diagonal fermions $\Psi \in \cH_{(u)} \otimes \cH_{(l)}^*$ for a background
of two branes
\begin{align}
X^a = \begin{pmatrix}
        X^a_{(u)} & \\ & X^a_{(l)}
       \end{pmatrix}.
\end{align}
Here $\Delta_a$ are the $SO(6)$ Gamma matrices.

As a warm-up,
consider first the intersection of a single $D0$--brane with $\R^2_\theta$ 
(cf. \cite{Berenstein:2012ts}). 
The $D0$ brane is given by a projector $X^a_{(u)} = p^a |p\rangle_{u}\langle p|_{u}$
 located at  $p^a = (p^4,p^5,p^6,0)$.
Then the fermions linking the state $|p\rangle_{u}$ with $\R^2_\theta$ have the form
\begin{align}
\Psi_p = |p\rangle_{u} \langle \psi|_{l}   
\label{left-localized-ansatz}
\end{align}
for some state $\psi$ on $\R^2_\theta$. Then we can write
\begin{align}
  \slashed{D}_{\rm int} \Psi &= \sum_{a=4,5,6} \Delta_a p^a \Psi -  \sum_{a=6,7} \Delta_a \Psi X^a_{(l)}  \nn\\
 &= \sum_{a=4,5} \Delta_a p^a \Psi - \Big(\Delta_6 \Psi (X^6_{(l)}-p_6) + \Delta_7 \Psi X^7_{(l)} \Big) \nn\\
 &=: \slashed{D}_{(1)} \Psi - \slashed{D}_{(2)} \Psi 
\end{align}
so that $\{\slashed{D}_{(1)},\slashed{D}_{(2)}\} =0$, and
$\slashed{D}^2_{\rm int} = \slashed{D}_{(1)}^2 + \slashed{D}_{(2)}^2$. Therefore $\Psi$ 
is a zero mode if and only if $ \slashed{D}_{(1)} \Psi =0= \slashed{D}_{(2)} \Psi$.
Clearly $\slashed{D}_{(1)} \Psi =0$ if and only if $p_4 = p_5 =0$ i.e. $p$ is located in 
$\cD_l$. Furthermore, it is easy to see following \cite{Chatzistavrakidis:2011gs}
that $\slashed{D}_{(2)} \Psi =0$ if and only if $|\psi\rangle_{l}$ is a coherent state on $\R^2_\theta$
localized at $p\in\cD_l$, with definite chirality associated with $\cD_l\cong\R^2$. 
This can be seen by introducing the shifted creation--and anihilation operators for $\R^2_\theta$
\bea
(X^6_{(l)}-p_6) + i X^7_{(l)} &=& a^\dagger, \qquad (X^6_{(l)}-p_6) - i X^7_{(l)} = a, 
\qquad [a,a^\dagger] = 2 \theta_{67} =: \theta
\label{y-oscillator}
\eea
which satisfy
\bea
(X^6_{(l)}-p_6)^2 + (X^7_{(l)})^2 &=& \frac 12(a^\dagger a + a a^\dagger) = \theta(\hat n + \frac 12) .
\eea
Here $\hat n = a^\dagger a$ is the   number operator.
We also introduce a fermionic oscillator representation for the Gamma matrices $\Delta_a$,
\begin{align}
2\a = \Delta_6 - i \Delta_7, \quad 2\a^\dagger =  \Delta_6 + i \Delta_7, 
\qquad  \{\a,\a^\dagger\} = 1. 
\end{align}
Hence the chirality operator on $\R^2_\theta$ is  given by 
\bea 
\chi \equiv \chi_{67} &=& i\Delta_6\Delta_7 = -2(\a^\dagger \a-\frac 12), 
\label{sub-chirality}
\eea
acting on the spin-$\frac 12$ irreducible representation. Moreover, it is straightforward to show that
\bea
\Sigma_{67} &=& \frac i4 [\Delta_6,\Delta_7] = \frac 12[\a,\a^\dagger] 
=  \frac 12\chi.
\eea 
With these tools, we can write
\begin{align}
 \slashed{D}_{(2)} \Psi &= \a \Psi a^\dagger + \a^\dagger \Psi a
   \label{Dirac-osc} \\
  \slashed{D}_{(2)}^2 \Psi &=\Psi  \theta(\hat n+\frac 12) - \Sigma \Psi \Theta_{(2)}  
 = \theta \Big(\Psi\hat n + \frac 12 (1+\chi) \Psi\Big) . 
\end{align}
From either equation it follows that $\slashed{D}_{(2)} \Psi =0$ if and only if 
$\Psi \hat n = 0 = (1+\chi)\Psi$,
which means that  $\psi_{(l)}$ is a coherent state on $\R^2_\theta$
localized at $p\in\R^2$, with definite helicity $\chi = -1$. 
Putting these results together, it follows that $\slashed{D}$ has zero modes linking the 
$D0$ brane with $\R^2_\theta$
\begin{align}
 \Psi_{p,s_i}^0 = |p,s_i\rangle_{u} \langle p,\downarrow|_{l} .
 \label{zeromode-ansatz-nohiggs}
\end{align}
It is remarkable that this is optimally localized at $p \in \R^2$. However,
there are two degenerate states with both chiralities $s_i=\pm 1$, 
corresponding to a vanishing index \cite{Steinacker:2013eya}.
If $p$ is located at some finite distance from $\R^2_\theta$, 
then these states are massive.

Now we switch on a Higgs field realized by the non-commutative $X^a_{(u)}$ 
given by the background \eq{branes-scheme-N2}. 
We denote the basis of $\cH_{(u)}$ with $|\pm\rangle_{u}$, so that 
\begin{align}
 X^6 |\pm\rangle_{u} = p_\pm^6 |\pm\rangle_{u} 
\end{align}
with
\begin{align}
 p_\pm^6 = \pm \frac r2 . 
\end{align}
We claim that now $\slashed{D}_{\rm int}$ has  precisely
one chiral zero mode located at each $p_\pm$. 
To see this, we write down again the Dirac operator for off-diagonal fermions 
acting on the above states as
\begin{align}
 \slashed{D}_{\rm int} \Psi  &= \slashed{D}_{(1)} \Psi - \slashed{D}_{(2)} \Psi
   \label{D-int-plane}
\end{align}
where now
 \begin{align}
  \slashed{D}_{(2)} \Psi &= \Delta_6\tilde X^6 \Psi + \Delta_7 \Psi X^7_{(l)} , \qquad 
 \tilde X^6\Psi = \Psi X^6_{(l)}- \frac r2 \s_3 \Psi
 \label{D-LR-def-Higgs}
\end{align}
  and
  \begin{align}
 \slashed{D}_{(1)} \Psi &= \frac{\phi}2(\s_+ \Delta_- + \s_- \Delta_+)\Psi, \qquad
 \Delta_\pm = \frac 12(\Delta_4 \pm i \Delta_5) .
  \end{align}
Noting that $\{\tilde X^6,X^7_{(l)}\}$ satisfy the same algebra $\R^2_\theta$
as  $\{X^6_{(l)},X^7_{(l)}\}$, we can introduce modified ladder operators
\begin{align}
 \tilde a \Psi &= \tilde X^6 \Psi - i \Psi X^7_{(l)}, \qquad 
  \tilde a^\dagger \Psi = \tilde X^6 \Psi + i \Psi X^7_{(l)},  \nn\\
  [\tilde a,\tilde a^\dagger] &= \theta
\end{align}
such that
\begin{align}
  \slashed{D}_{(2)} \Psi &= \a \tilde a^\dagger \Psi  + \a^\dagger\tilde a \Psi 
   \label{Dirac-osc-mod} 
\end{align}
and therefore
\begin{align}
 \slashed{D}_{(2)}^2 \Psi &= \theta_a(\tilde a\tilde a^\dagger +\frac 12)\Psi  - \Sigma \Psi \Theta_{(l)} 
= \theta \Big( \tilde a\tilde a^\dagger \Psi+ \frac 12 (1+\chi) \Psi\Big) .
\end{align}
Therefore $\slashed{D}_{(2)} \Psi = 0$  if and only if $ \tilde a\tilde a^\dagger \Psi = 0 = (1+\chi) \Psi$. 
This is equivalent to 
\begin{align}
 \tilde a^\dagger \Psi= - \frac r2\s_3 \Psi + \Psi (X^6_{(l)} +i X^7_{(l)}) = 0=  \tilde \a^\dagger \Psi ,
\end{align}
which means that $\Psi$ consists of coherent states 
localized at $X^6=\frac r2 \s_3$ and $X^{7} = 0$. More explicitly, expanding $\Psi$ in the appropriate
basis $\{|+\rangle_{u} \langle n_+|_{l}, |-\rangle_{u} \langle n_-|_{l}\}$ (dropping helicities)
where $\langle n_\pm|_{l}$ denotes an oscillator basis with origin $p_\pm$,
it follows that the zero modes of $\slashed{D}_{(2)}$ have the form 
\begin{align}
\Psi_{+(2)} &= |+,s_+\rangle_{u} \langle p_+,\downarrow|_{l},  \nn\\
 \Psi_{-(2)} &= |-,s_-\rangle_{u} \langle p_-,\downarrow|_{l}   
\end{align}
Imposing in addition $\slashed{D}_{(1)} \Psi = 0$ using \eq{D-LR-def-Higgs}
and noting that $\Delta_+|+,\uparrow\rangle_{u} =0$ and $\Delta_-|-,\downarrow\rangle_{u} =0$ 
leaves the following two zero modes of $\slashed{D}_{\rm int}$
\begin{align}
\Psi_{+L} &= |+,\uparrow\rangle_{u} \langle p_+,\downarrow|_{l},  \nn\\
 \Psi_{-R} &= |-,\downarrow\rangle_{u} \langle p_-,\downarrow|_{l}   
 \label{zeromodes-Higgs-N2}
\end{align}
with definite chirality $L,R$ in $\R^4$.
These are  the two chiral zero modes located at $p_\pm$ 
expected from the picture of intersecting branes.
As a check, it is straightforward to verify using \eq{Dirac-osc-mod} that  the states \eq{zeromodes-Higgs-N2}
are indeed zero modes of $\slashed{D}_{\rm int}$.

To summarize, we have found that switching on $\phi$, i.e., fusing the two points to a minimal fuzzy ellipsoid, 
lifts the degeneracy of the two polarization states at a single point. One is then left with two zero modes of 
opposite chirality, located at the opposite poles of the fuzzy ellipsoid.

If $\cD_l$ is described by some curved brane, then these would-be zero modes 
acquire some small mass.
The associated Yukawa coupling will be proportional to the Higgs $\phi$, as discussed in the next
section.

\subsection{$\R^2 \times S^2$ intersecting $\R^2$}
\label{sec:intersecting-plane-4D}
 
 Finally we add the missing $\R^2_{\theta (89)}$ to $\cD_{u} = \R^2_{89} \times S^2$.
 This is achieved simply by adding $\Delta_8 X^8 + \Delta_9 X^9$ to  $\slashed{D}_{(\rm int)}$ and $\slashed{D}_{(1)}$,
 leading to an additional term
 \begin{align}
 \b b^\dagger\Psi  + \b^\dagger b\Psi  
 \end{align}
 where $b,b^\dagger$ form the oscillator representation of $\R^2_{\theta (89)}$ and 
 \begin{align}
2\b = \Delta_8 - i \Delta_9, \quad 2\b^\dagger =  \Delta_8 + i \Delta_9, 
 \qquad  \{\b,\b^\dagger\} = 1 . 
 \label{beta-def}
\end{align}
The additional contribution vanishes if and only if $b\Psi =0 = \b \Psi$, so that 
the above results generalize immediately. We obtain 
the following two zero modes of $\slashed{D}_{\rm int}$
\begin{align}
\Psi_{+L} &= |+0,\uparrow\downarrow\rangle_{u} \langle p_+,\downarrow|_{l},  \nn\\
 \Psi_{-R} &= |-0,\downarrow\downarrow\rangle_{u} \langle p_-,\downarrow|_{l}   
 \label{zeromodes-Higgs-N2-2}
\end{align}
with definite chirality $L,R$ in $\R^6$.
These are  the two chiral zero modes located at $p_\pm$ 
expected from the picture of intersecting branes.

Finally, note that 
the Dirac equation for these fermionic would-be zero modes are not affected by $H^a_{(S)}$ 
due to  the coherent state property $h^a X_a |0\rangle_u =0$,
except for $\nu_R$, which could acquire a Majorana mass term via the 
gauge singlet $\tr_N(\nu_R S)$.

 \paragraph{Mirror fermions.}
 
 Besides these zero modes, there are additional pairs of massive fermions (``mirror fermions'')
 with opposite chirality at the same intersections, coupling to the same gauge fields.
 The lowest ones  arise from the opposite helicity of the coherent state $|+\rangle$
on the minimal $S^2_2$, with eigenvalue of $\slashed{D}_{(1)}$ of order $2\phi$.
We denote those by $\widetilde\psi$.
Additional sets of ultra-massive fermions with mass of order $\theta$ arise from 
other helicity and oscillator states on $\R^2_\theta$.
In this way, a chiral model emerges at low energies from the non-chiral underlying $\cN=4$ theory,
 with a large hierarchy between the low-energy chiral fermions and their massive mirror partners.
 Such a mirror model could be phenomenologically viable provided the hierarchy is sufficiently large.

 One potential problem  is the fact that the tree level mass of 
 these lowest mirror fermions is only by a factor $\sqrt{2}$ higher than the tree-level $W$ mass, 
 both being determined by $\phi$ (see section \ref{sec:4D-masses}).
 However, this is also the scale of the KK modes on the large branes,
 which couple to the fermions but not to the electroweak gauge fields.
 It then seems reasonable that quantum effects increase the mass of the mirror fermions sufficiently high above
the electroweak scale.

\subsection{Deformations, would-be zero modes and Yukawa couplings}
\label{sec:Yukawas}

\subsubsection{Analytical expectation}

Armed with these results for the flat case, we would like to understand the fermions  arising at the 
intersections of the compact branes $\cD_u\cap \cD_l$ given by large and small fuzzy spheres \eq{branes-explicit}.
We expect that the qualitative features of the flat limit survive: there should be pairs of 
near-zero eigenmodes of $\slashed D_{\rm int}$ called would-be zero modes henceforth, 
which due to $\slashed D_{\rm int}\Gamma^{(\rm int)} = - \Gamma^{(\rm int)} \slashed D_{\rm int}$ 
(cf. \eq{Gamma-int}) decompose into
states $\Psi_L, \Psi_R$ of definite chirality, which in turn are approximately localized at the 
intersections $p_\pm$ of the branes. 
However, the helicities should be determined by the local tangent planes
at the intersections. 
We therefore  expect that the following ansatz in terms of coherent states should be appropriate
\begin{align}
\Psi_{+L} &= |+0,\nwarrow\searrow\rangle_{u} \langle p_+,\searrow|_{l} ,  \nn\\
 \Psi_{-R} &= |-0,\swarrow\swarrow\rangle_{u} \langle p_-,\swarrow|_{l}  , 
 \label{zeromodes-Higgs-deformed}
\end{align}
at least if the intersection\footnote{Here the classical orbit of the loci of the coherent states on the fuzzy 
branes is relevant. For the fuzzy sphere, this has radius $R \frac {N-1}2$ instead of $R \frac{\sqrt{N^2-1}}2$.}
is perpendicular. Here the coherent states are located at the intersection
of the branes, with slightly modified helicity orientation reflecting the local geometry.
The incompatible spin orientations of the pair of would-be zero modes
leads to non-vanishing Yukawa couplings and eigenvalues. 
To gain some analytic insights,
let us compute these Yukawa couplings explicitly for the above ansatz \eq{zeromodes-Higgs-deformed}.
Consider first
\begin{align}
 \tr_N \Psi_{-R}^* \g_0\g_5 \slashed{D}_{\rm int} \Psi_{+L}
  &= \tr_N \Big(|p_-,\swarrow\rangle_{l} \langle -0,\swarrow\swarrow|_{u}  \g_0\g_5 \slashed{D}_{\rm int} 
     |+0,\nwarrow\searrow\rangle_{u} \langle p_+,\searrow|_{l}\Big) \nn\\
 &= \frac 12\langle p_+,\searrow |p_-,\swarrow \rangle_{l} \langle -0,\swarrow\swarrow|\phi(\s_+ \Delta_- + \s_- \Delta_+)|+0,\nwarrow\searrow\rangle_{u}  \nn\\
 &\approx  \phi\, 
   \langle p_+,\searrow  | p_-,\swarrow\rangle_{l} \langle -0,\swarrow\swarrow|\Delta_+|-0,\nwarrow\searrow\rangle_{u} \nn\\
   &=  \phi f_{RL}
  \label{updown-yukawa} 
 \end{align}
In the last step, we observed that only the second term in 
$(\s_+ \Delta_- + \s_- \Delta_+)$ can give non-vanishing matrix elements
between $\langle-,\swarrow|_{u}$ and $|+,\nwarrow\rangle_{u}$, and evaluated the action of $\s_-$.
The contribution from the coherent states on $\cD_{l}$ can be approximated by
\begin{align}
 \langle p_+,\searrow | p_-,\swarrow\rangle_{l} 
 &\approx  \langle p_+ | p_-\rangle_{l}
\end{align}
since the two spin directions should be appropriately aligned, as long as $\cD_{l}$ is much larger than $S^2_2$.
In the flat limit this inner product would be exponentially suppressed 
with the distance of the two coherent states on $\cD_{l}$,
\begin{align}
 \langle p_+ | p_-\rangle_{l} 
 \approx e^{-\frac{(p_- - p_+)^2}{R^2N}}
\end{align}
although this factor is typically very close to 1 for the compact branes under consideration.
On the other hand, the  contribution from the coherent states on $\cD_u$ can be approximated by
the spin contribution only,
\begin{align}
\langle -0,\swarrow\swarrow|\Delta_+|-0,\nwarrow\searrow\rangle_{u} 
  &= \langle \swarrow\swarrow|\Delta_+|\nwarrow\searrow\rangle_{u} .
  \label{sphere-gamma-inner}
\end{align}
This is non-vanishing only due to the non-alignment of the two helicities at the two intersections. Assuming that 
the spinor wavefunctions factorize (as in the flat case), the spinor associated with $S^2_2$ is given by 
\begin{align}
\label{eq:HelicityAnsatz}
 |\nwarrow\rangle_{u} = \begin{pmatrix}
                           1-\frac 12\e^2 \\
                           \e
                          \end{pmatrix}, \qquad
|\searrow\rangle_{u} = \begin{pmatrix}
                           \e \\
                           1-\frac 12\e^2
                          \end{pmatrix}, 
\end{align}
where $\e \sim \sin(\frac 12\vartheta)$ and $\vartheta$ measures the angle of  the two distinct 
tangent planes $T_{p_\pm}\cM$ relative to the flat limit.
This  gives
\begin{align}
 \langle\swarrow|\Delta_+ |\nwarrow\rangle_{u}
  & \ \approx  \ \e^2   \ \approx \ \frac{r^2}{4N_l^2R_l^2}
\end{align}
using \eq{intersect-class} for the present geometry.
Combining these results, we obtain the desired Yukawa couplings
\begin{align} 
\label{eq:f_RL}
f_{RL} \approx \, \frac{r^2}{4N_l^2R_l^2}\,  e^{-\frac{r^2}{NR^2}} .
\end{align}
This is clearly small for the would-be zero modes under consideration, 
while for the mirror fermions $\widetilde \psi$ with reversed spin associated with the $S^2_2$ 
we would get 
\begin{align}
\widetilde f_{LR} \ \approx \  \, e^{-\frac{r^2}{NR^2}}
\end{align}
due to $\langle \nearrow|\Delta_+|\searrow\rangle_{u} \approx 1$.
In particular, there is naturally a large hierarchy $\cO(\e^2)$ between the lowest chiral sector corresponding
 to the standard model, and the first series of massive mirror fermions.
 Moreover these quantities are accessible, both to (refined) analytical 
considerations and to numerical methods.   
Of course they will also be subject to quantum corrections, which are out of the scope of the present paper.
One may hope that these quantum corrections help to increase the separation of the 
mirror fermions from the electroweak $W,Z$ bosons, as discussed below.

\subsubsection{Numerical results}
\label{sec:numerics}

\begin{figure}
 \includegraphics[width=0.85\textwidth]{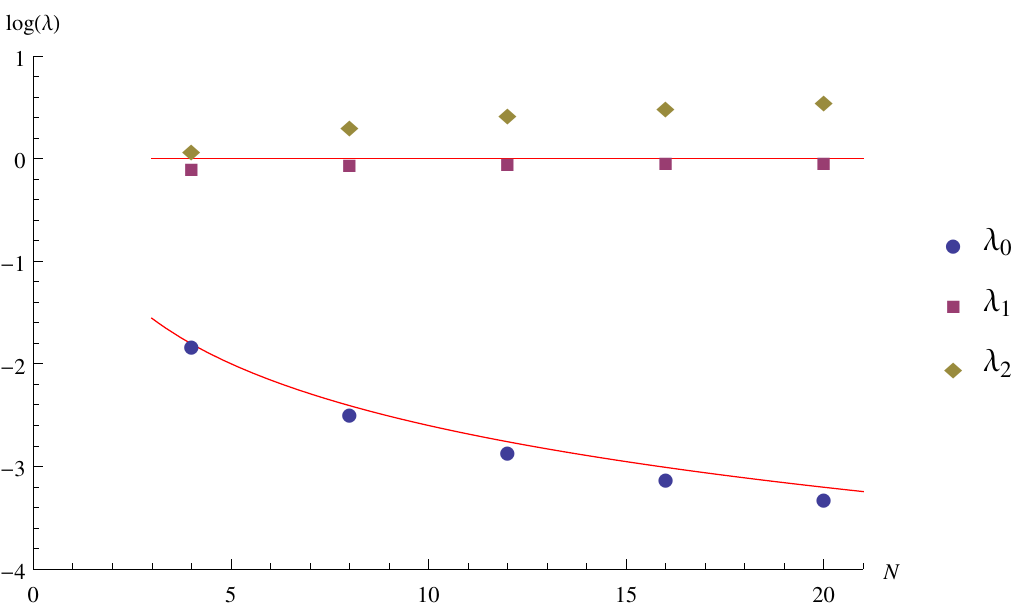}
 \caption{Lowest eigenvalues as a function of $N$, for $N_{l,u} = N$, $R_{l,u} = R'_{l,u} = 1$, and $r = \phi = 1$, with the theoretical expectations \eqref{eq:lambda_0} and \eqref{eq:lambda_1}.}
 \label{fig:PlotA}
\end{figure}

\begin{figure}
 \includegraphics[width=0.85\textwidth]{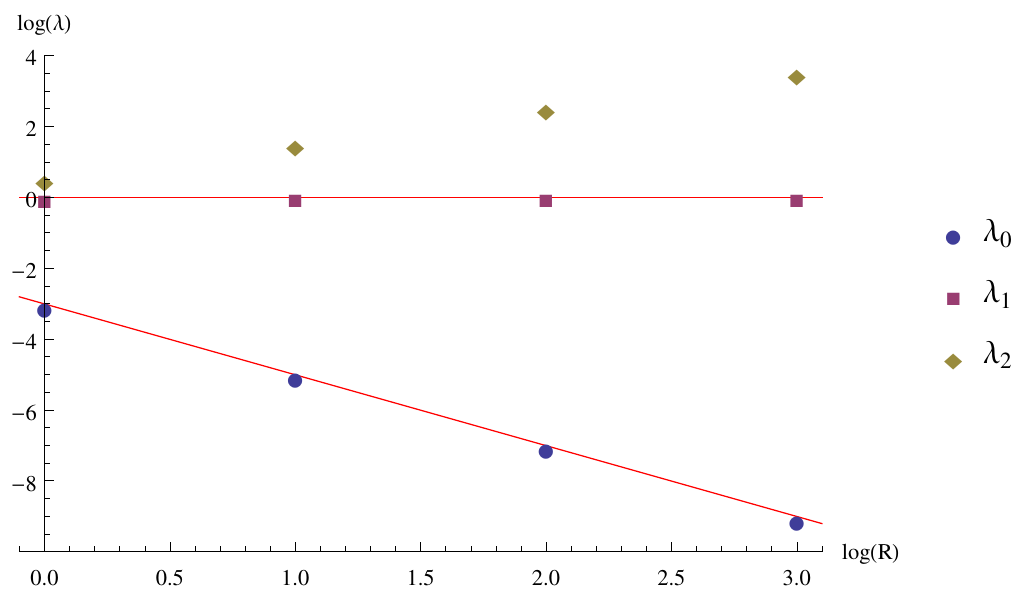}
 \caption{Lowest eigenvalues as a function of $R$, for $N_{l,u} = 16$, $R_{l,u} = R'_{l,u} = R$, and $r = \phi = 1$, with the theoretical expectations \eqref{eq:lambda_0} and \eqref{eq:lambda_1}.}
 \label{fig:PlotB}
\end{figure}

\begin{figure}
 \includegraphics[width=0.85\textwidth]{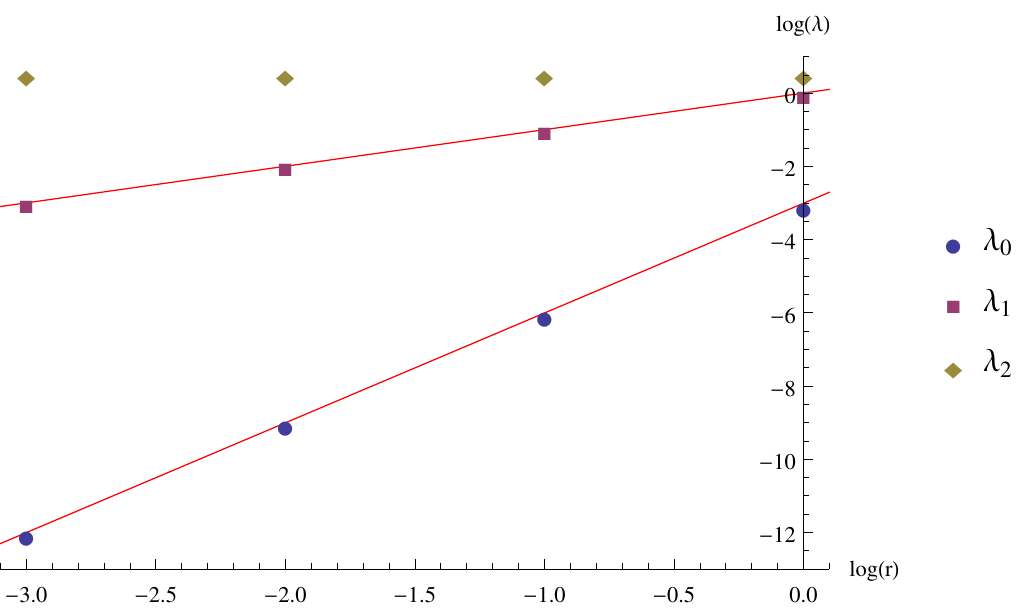}
 \caption{Lowest eigenvalues as a function of $r$, for $N_{l,u} = 16$, $R_{l,u} = R'_{l,u} =1$, and $r = \phi$, with the theoretical expectations \eqref{eq:lambda_0} and \eqref{eq:lambda_1}.}
 \label{fig:PlotC}
\end{figure}

Some aspects of the theoretical expectations derived in the previous subsection can be verified numerically.
We compute the eigenvalues of the internal Dirac operator for the off-diagonal spinors 
connecting the two branes, and identify them with the Yukawa couplings of their
chiral components.
Restricting to a regime where the Planck cells of the larger fuzzy spheres are greater than $r$, 
we can neglect the Gaussian factor in \eqref{eq:f_RL}, and obtain for the lowest eigenvalue
\begin{equation}
\label{eq:lambda_0}
 \lambda_0 \approx \frac{\phi r^2}{4 N_l^2 R_l^2},
\end{equation}
and for the next eigenvalue, i.e., the mirror fermions,
\begin{equation}
\label{eq:lambda_1}
 \lambda_1 \approx  \phi.
\end{equation}
In Figures~\ref{fig:PlotA}, \ref{fig:PlotB}, and \ref{fig:PlotC}, 
we see that these estimates agree quite well with numerical results (the red lines show the expectations from \eqref{eq:lambda_0} and \eqref{eq:lambda_1}). Also the next eigenvalue $\lambda_2$ is shown. Furthermore, we see that one can produce large hierarchies for moderate $N$.

\begin{figure}
 \includegraphics[width=0.85\textwidth]{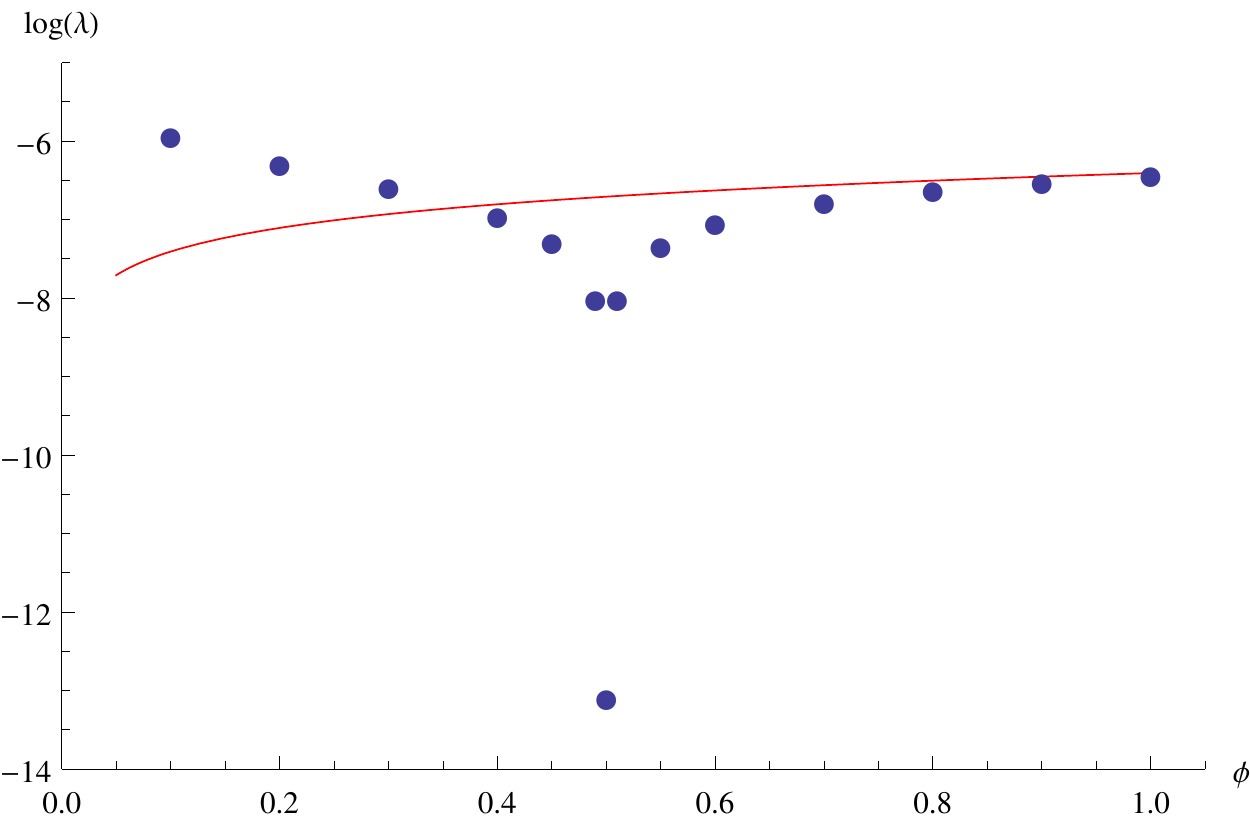}
 \caption{Lowest eigenvalue as a function of $\phi$, for $N_{l,u} = 8$, $R_{l,u} = R'_{l,u} =1$, and $r = 1$, with the theoretical expectation \eqref{eq:lambda_0}.}
 \label{fig:PlotD}
\end{figure}

However, varying $\phi$ while keeping $r$ fixed leads to a dramatic deviation from the expectation \eqref{eq:lambda_0}, 
as shown in Figure~\ref{fig:PlotD} (note that the scale is logarithmic): There is a very pronounced minimum of 
$\lambda_0$ at $\phi = r/2$ (which is excluded by the equation of 
motion in Section~\ref{sec:intersecting-brane-solutions}). This is also seen for other choices of $N, R$, so it seems to be a 
universal behavior. From geometrical considerations, one would rather expect $\phi \approx \frac{1}{\sqrt{2}} r$ to be 
special, as then the branes intersect orthogonally. Hence, a complete understanding of the Yukawa couplings is missing, 
but the generation of a large gap between the lowest and the next eigenvalue of $\slashed{D}_{\rm int}$ is certainly possible.

\begin{figure}
 \includegraphics[width=0.85\textwidth]{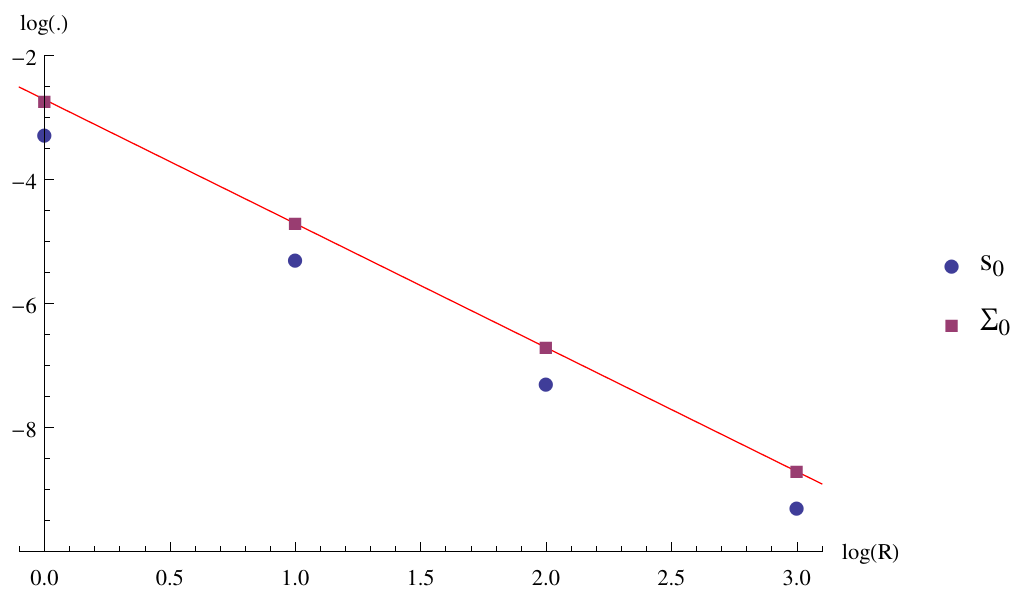}
 \caption{Expectation values of $s = 1-\sigma_3$ and $\Sigma=1-\Sigma_{45}$ in the lowest eigenvalue as a function of $R$, for $N_{l,u} = 16$, $R_{l,u} = R'_{l,u} = R$, and $r = \phi = 1$, with the theoretical expectation \eqref{eq:SigmaPrediction}.}
 \label{fig:PlotB2}
\end{figure}

For our arguments, it is crucial that the lowest eigenstates, when projected to a definite 6-dimensional chirality, 
are very well localized at $\pm (0,0,1)$ on $S^2_2$, and are essentially eigenvectors of $\chi_{45} = 2  \Sigma_{45}$, 
the chirality operator corresponding to the $4-5$ plane.\footnote{Note that they are then also essentially eigenvectors of 
$\chi_{67} \chi_{89}$. Projecting on the eigenspaces of, say, $\chi_{89}$, one obtains a vector which is essentially an 
eigenvector of $\Sigma_{45}$, $\Sigma_{67}$, and $\Sigma_{89}$.} From \eqref{eq:HelicityAnsatz}, we expect the expectation 
value of $1-\Sigma_{45}$ in the lowest eigenstate $\psi_0$ to be roughly
\begin{equation}
\label{eq:SigmaPrediction}
 \bra{\psi_0} (1-\Sigma_{45}) \ket{\psi_0} \approx 2 \epsilon^2 \approx \frac{r^2}{2 N_l^2 R_l^2}.
\end{equation}
Figure~\ref{fig:PlotB2} shows the expectation value of $1-\sigma_3$ and $1-\Sigma_{45}$ in the lowest eigenstate as a 
function of $R=R_{l,u} = R'_{l,u}$. Also the expectation \eqref{eq:SigmaPrediction} is 
plotted.\footnote{Note that our ansatz was that $\psi_0$ is an eigenstate of $\sigma_3$ corresponding to $S^2_2$, so the 
above discussion does not give a prediction for the expectation value of $1-\sigma_3$.} We see that the deviation from 
being an eigenstate indeed decreases for increasing $R$.

 \subsection{Gauginos}

 Besides the fermions arising at the interactions of the various branes, 
 fermions also arise in the diagonal blocks, as functions on the corresponding branes. 
 They can be viewed as (generalized) gauginos, i.e. superpartners of the gauge bosons 
 or scalar fields.
 All these fermions are non-chiral, i.e. both chiral sectors couple identically to the gauge 
 and scalar fields. 
  This includes the  gluinos, binos, winos, Higgsinos, etc.
 Note that the gauginos corresponding to $U(1)_B$ and $U(1)_l$ are neutral 
 under the full standard model gauge group, and therefore decouple
 at low energies. As usual they may be considered as dark matter candidates.
 The gluinos and other gauginos are expected to get radiative mass.
 Furthermore, there are towers of higher KK modes for all these fermions.
 However, note that the present backgrounds are far from supersymmetric, since e.g. the scalar superpartners 
 of the standard model fermions have tree level mass of order $\theta \sim R^2 N$.

\subsection{Fermion masses and Yukawas}

In this section, we show that the 4-dimensional masses of the would-be zero modes 
are given by the corresponding Yukawa couplings.
As recalled in Appendix  \ref{sec:clifford}, \cf also \eqref{S-YM-eff},
the Dirac operator can be written as 
\begin{align}
 \slashed{D} \Psi &= \left( \slashed{D}_4 + \rho^{\frac{1}{2}} \g_5 \slashed{D}_{\rm int} \right) \Psi.
\end{align}
Here
\begin{align}
 \slashed{D}_{4} &= \tilde \gamma^\mu \left( i\del_\mu + [\cA_\mu, \cdot ] \right),
\end{align}
is the massless 4-dimensional Dirac operator on $\R^4_\theta$ with 
$\tilde \gamma^\mu = \rho^{\frac 12} \theta^{\nu\mu}\gamma_\nu$,
and $\slashed{D}_{\rm int}$ is defined in terms of the $X^a$.
Now consider a pair of eigenspinors $\psi_{\pm}$ of 
the internal Dirac operator
\begin{align}
 \rho^{\frac 12} \slashed{D}_{\rm int}\psi_{\pm} = \pm m \psi_{\pm}, \qquad 
 \psi_{\pm} = \psi_\uparrow \pm \psi_\downarrow  
\end{align}
in terms of two chirality eigenstates (such as our would-be zero modes),
with $\Gamma^{\rm (int)} \psi_\updownarrow = \pm  \psi_\updownarrow$ to be specific. 
Then $ \rho^{\frac 12} \slashed{D}_{\rm int}\psi_\uparrow = m \psi_\downarrow$, corresponding to a Yukawa coupling $m$.
Now consider a 32-component spinor whose internal components consist of the above two internal helicity states,
\begin{align}
\Psi =  \chi_1\otimes\psi_\uparrow + \chi_2\otimes\psi_\downarrow
\end{align}
where $\chi_{1,2}$ are Dirac spinors of $SO(3,1)$.
This can be represented as an auxiliary 8-component spinor consisting of the two Dirac spinors of $SO(3,1)$ only,
\begin{align}
 \Psi \cong \begin{pmatrix}
          \chi_1 \\   \chi_2
        \end{pmatrix} 
 \label{spinors-p}
\end{align}
and the 10-dimensional Dirac equation can be written as
\begin{align}
0=  \slashed{D} \Psi &= 
 \begin{pmatrix}
  \slashed{D}_4 &  \g_5 m  \\
   \g_5 m  & \slashed{D}_4
 \end{pmatrix}\Psi \ .
 \label{Dirac-fulll}
 \end{align}
This has solutions with 4-dimensional mass $m$, since
\begin{align}
 \slashed{D}^2 &=  \begin{pmatrix}
  \slashed{D}_4^2 + m^2 & 0 \\
  0 & \slashed{D}_4^2 + m^2
 \end{pmatrix} 
\end{align}
noting that $\{\g^5,\tilde \g^\mu\} = 0$.

Finally we show that the scalar Higgs $S$ \eq{S-ansatz} does not affect the fermionic would-be zero modes.
It contributes to the internal Dirac operator as follows
\begin{align}
 \slashed{D}_{\rm int}^{(S)} \psi &= \Delta_a[X^a_{(S)},\psi] = 
  2  h \beta^+[S,\psi] + 2  h \beta [S^\dagger,\psi] 
\end{align}
recalling the definition of $\beta$ in section \ref{sec:intersecting-plane-4D}. This vanishes due to 
$\b \psi = 0$ for the would-be zero modes, using the explicit form  of $S$.

\section{Symmetry breaking and  4-dimensional fields}
\label{sec:gauge-fields}

Spelling out all the branes, the background \eq{branes-explicit} is given by
\begin{align}
 X^4 + i X^5 &= \begin{pmatrix}
             R'_u L_3 &0 & 0 & \phi \one& &\\
             0& R'_d L_3 & \phi \one& 0&&\\
             0& 0& R'_d L_3  & 0&&\\
             0 & 0& 0 & R'_u L_3  &&\\
             &&&& R'_2 K_3&\\
             &&&&& R'_2 K_3 \one_3
            \end{pmatrix}  ,  \nn\\[1ex]
 X^6 + i X^7 &= \begin{pmatrix}
               -\frac r2 \one &&&&&\\
               &- \frac r2 \one&&&&&\\
               && \frac r2\one &&&\\
               &&& \frac r2\one &&\\
               &&&& R K_+&\\
               &&&&& R K_+ \one_3
              \end{pmatrix},  \nn\\[1ex]
 X^8 + i X^9 &=  \begin{pmatrix}
            R L_+ & & & &&\\
             & R L_+ &&&& \\
              & & R L_+ &&&& \\
               & & & R L_+ & 2h S &&\\
                &&&&0&\\
                &&&&&&0_3
            \end{pmatrix} .
\label{EW-branes}
\end{align}
We also included  the Higgs singlet $S$,
which connects the branes $\cD_u$ and $\cD_l$. 

We want to understand the bosonic modes which arise as fluctuations
$X^a \to X^a + \cA^a$
about the above background. 
In general, such fluctuations on a fuzzy brane $\cK_N$ can be written as a finite sum
\begin{align}
 \cA(x,y) = \sum_{lm} \cA_{lm}(x) Y^{lm}
\end{align}
where $Y^{lm}\in Mat(N,\C)$ stands symbolically for the harmonics of 
$\Box = [X^a,[X^a,.]]$ on $\cK_N$. 
This applies both to scalar fields, gauge fields and the gauginos.
It  provides a geometric interpretation of the matrix-valued  fields on $\R^4$ in terms of 
towers of massive Kaluza-Klein modes  on $\cK_N$. 
On the fuzzy sphere $S^2_{N}$, this KK tower arises at roughly equidistant masses determined by the eigenvalues of $\Box$,
with lowest non-trivial eigenvalue $\sim R^2$. Therefore 
at low energies, it suffices to keep only the massless modes $\sim\one$ on the $S^2_{N}$.
Then the $\cA(x,y)$ can be viewed as functions on $\R^4$ taking values in the above space of $8\times 8$ matrices.
In particular, the stack of three coincident $\cD_B$ branes gives rise to the massless gluons 
with unbroken $U(3) = SU(3)_c \times U(1)_B$ gauge symmetry, as well as an
associated finite tower of massive KK modes.
On the other hand, the KK tower on the 
minimal branes is very short, and contains in particular the electroweak Higgs, 
the $Z$ boson, and the $B_5$ and $C_\mu$ bosons as discussed below.

\subsection{Hierarchical symmetry breaking}

To determine the masses of the low-energy gauge bosons explicitly,
it is useful to first replace the two minimal fuzzy spheres $S^2_{N=2}$ by a stack of 4 coincident D0 branes 
$X^a_{(1)} = p^a \sum_{i=1}^4|i\rangle\langle i|$ located at 
some point $p^a$ on the coincident $\cD_l\cong \cD_B$ branes 
described by $\one_4 \otimes X^a_{(2)}$.
This background admits a $U(4) \times U(4)$ symmetry.
Now we switch on a non-vanishing singlet Higgs $S \sim |i\rangle\langle p|_{l}$, 
where $|p\rangle_{l}$ is a coherent state on $\cD_{(l)}$ located at the  $D0$ branes $p^a$.
Since $S$ has rank one, it breaks the  
symmetry to the commutant $U(3)_B \times U(3) \times U(1)_S$, where
$U(1)_S$ acts diagonally on $|i\rangle \oplus |p\rangle_{l}$.
This $U(1)_S$ can be traded for $U(1)_{\rm tr}$, which has a clear geometric interpretation.
We assume here that this breaking happens at 
a high scale, and restrict ourselves to the commutant of $S$ from now on. 
The fermionic modes on such a background are still non-chiral.

Next, we introduce the long axis along $X^6$ of the electroweak ellipsoids by turning on $r>0$.
This breaks the above symmetry further to the commutant of $X^6$ (in $U(3)_B \times U(3) \times U(1)_{\rm tr}$). 
The bosons $C_\mu$ associated with this breaking will be discussed below.
Using the explicit form \eq{EW-branes}, this commutant  is 
given by $U(3)_B \times SU(2)_L \times U(1)_Y \times U(1)_5\times U(1)_{\rm tr}$,
with generators\footnote{Here $I_{N_1}$ indicates the identification $\C^{N_u} \cong \C^{N_d}$ of the Hilbert spaces 
of $S^2_{N_u}$ and $S^2_{N_d}$, assuming $N_u = N_d = N_1$.}
\begin{align}
 t_{\pm,3} &= \frac 12\begin{pmatrix}
              \sigma_{\pm,3} I_{N_1}& & &&\\
             & 0 & &&\\
             & & 0 &&\\
             &&& 0 &\\
             &&&& 0
            \end{pmatrix}, \qquad
 t_Y =  \begin{pmatrix}
             0_2 & & && \\
             & -\one_{N_1}  &  && \\
             & & \one_{N_1} && \\
             &&& \one_{N_2} & \\
             &&&& -\frac 13\one_{N_2}
            \end{pmatrix},  \nn\\[1ex]
 t_5 &= \begin{pmatrix}
         \one_2\otimes\one_{N_1} &&&&\\
          & -\one_{N_1} &&&\\
          && -\one_{N_1} & &\\
          &&&-\one_{N_2} & \\
          &&&& \frac 13\one_{N_2}
        \end{pmatrix}  , \qquad
 t_{(s)\a} = \begin{pmatrix}
          0_2 &&&&\\
          & 0 &&&\\
          && 0& &\\
          &&& 0 & \\
          &&&& \l_\a \one_{N_2}
        \end{pmatrix}   
 \label{standard-generators}
\end{align}
Here we assume that $\cD_{l,B}$ is represented on $\C^{N_2}$,
and $\l_\a \in \mmu(3) = \msu(3) \oplus \mmu(1)_B$.
Note that $t_5$ acts as 
\begin{align}
 [t_5,.] \cong B-l+\g_5
\end{align}
on the fermionic zero modes. It is therefore anomalous and expected to 
disappear from the low-energy spectrum, along with $U(1)_B$. 
This leaves exactly the gauge group of the standard model 
$SU(3)_c \times SU(2)_L \times U(1)_Y$, extended by the anomalous $U(1)_B, U(1)_5$, 
and the geometric $U(1)_{\rm tr}$.
The $U(1)_5$ is also broken by the electroweak Higgs, as elaborated below.

Finally, we switch on the electroweak Higgs $\phi_{u,d}$, so that the 4 $D0$ branes 
expand to form two fuzzy ellipsoids $S^2_2$.
Then the symmetry breaks down as desired to  
$SU(3)_c \times U(1)_Q \times U(1)_B\times U(1)_{\rm tr}$, with charge generator 
\begin{align}
 t_Q &=  t_3 + \frac 12 t_Y = \frac 12(\one_u - \one_d + \one_l - \frac 13 \one_B) .
 \label{el-charge}
\end{align}
Here $U(1)_B$ is anomalous,
and  $U(1)_{\rm tr}$ is a geometric mode associated to gravity.

As a check, 
the unbroken gauge group of the above brane configuration can alternatively be obtained 
as follows.
Consider first the background without $S$, given by 
two coincident branes  $\cD_u \simeq \cD_d$  and 
4 coincident branes $\cD_l \simeq \cD_B$.
This background has an unbroken $U(2) \times U(4)$ gauge symmetry. 
Now we switch on the  scalar Higgs $S$. This breaks the symmetry to its commutant
$SU(3)_c \times U(1)_Q \times U(1)_B\times U(1)_{\rm tr}$.

\subsection{Four-dimensional gauge bosons and masses}
\label{sec:4D-masses}

We recall the 4-dimensional form of the effective action \eq{S-YM-eff}
of the matrix model in our brane background.
To obtain the proper coupling constants for the corresponding gauge fields, we
introduce canonically normalized generators with $\tr_N (\tilde t_i \tilde t_j) = \frac 12 \d_{ij}$, via
\begin{align}
 t_{\pm 3}  &= c_L \tilde t_{\pm,3}, \qquad  c_L^2 = N_1  \nn\\ 
 t_Y  &= c_Y \tilde t_Y, \qquad  c_Y^2 = 2(2N_1 + \frac 43 N_2)  \nn\\ 
 t_5  &= c_5 \tilde t_5, \qquad  c_5^2 = 2(4N_1 +\frac 43 N_2)  \nn\\ 
 t_\a &= c_S \tilde t_\a,\qquad  c_S^2 = N_2 \ .
\end{align}
The point is that the $\tilde t_\a$ act on the full Hilbert space of the matrix model 
and satisfy rescaled commutation relations, 
while the $t_\a$ act on the reduced Hilbert space of physical fermionic states as in the standard model. 
To identify the low-energy gauge couplings, we write 
the gauge fields in two ways using \eq{standard-generators} 
 \begin{align} 
  \cA &= g_{\rm YM} \Big( W_- \tilde t_+ +  W_+ \tilde t_- +  W_3 \tilde t_3 
  + B \tilde t_Y +  B_5 \tilde t_5 +  A_\a \tilde t_\a \Big) \nn\\
   &= g ( W_- t_+ +  W_+ t_- + W_3 t_3) 
   + \frac 12 g' B\, t_Y  +  g_5 B_5\, t_5 + g_S A_\a  t_\a 
 \end{align}
 where
 \begin{align}
g =  \frac{g_{\rm YM}}{c_L}, \quad 
 g_S = \frac{g_{\rm YM}}{c_S}, \quad 
 \frac 12 g' = \frac{g_{\rm YM}}{c_Y},
 \quad  g_5 = \frac{g_{\rm YM}}{c_5} .
\end{align}
The effective standard model coupling constants in the second line are identified
from the covariant derivative on the fermions
\begin{align}
 i D_\mu \psi = \theta^{-1}_{\mu\nu}[X^\nu,\psi] 
  &= \big(i\del_\mu + [\cA_\mu,.]\big) \psi  \nn\\
  &\stackrel{!}{=} \big(i\del_\mu + g W_i t_i + \frac{g'}2 B t_Y + g_S A_\a t_\a +  g_5 B_5\, t_5\big) \psi 
 \end{align}
 on the fermionic would-be zero modes. 
 Since the relevant fermionic (would-be) zero modes are 
 made of one-dimensional (coherent) states in the internal Hilbert space, 
 the term $\tr_N\bar\psi\tilde\g^{\mu}iD_\mu\psi$ in \eq{S-YM-eff}
 reduces to the appropriate  Lagrangian for the 4-dimensional
 fermions in the standard model, without any extra factors coming from $\tr_N$.
 We can therefore identify 
 the gauge fields $W_i, B$, etc.  with those of the the standard model, where
$g$ is the  $SU(2)_L$ coupling constant, $g'$ is the $U(1)_Y$ coupling constant,
 $g_S$ is the strong coupling constant, and $g_5$ the one associated with $U(1)_5$. 
 These tree-level couplings apply at very high energies. 
 The kinetic terms of these gauge fields have the standard normalization, 
 and by gauge invariance their full action must be 
 \begin{align}
  S_{\rm YM} &= -\int d^4 x\, \frac 1{4g_{\rm YM}^2} \tr_N (\cF \cF)_G 
   = -\int d^4 x\, \frac 1{4} \tr_{\rm red}(F F)_G + ...  
 \label{gaugefields-SM}
\end{align}
Here $\tr_{\rm red}$  is the trace in the adjoint representation of 
the reduced low-energy gauge group $\oplus_i \mg_i$ generated by the $t_i$,
with gauge fields $A,W,$ etc. corresponding to the (extended) standard model;
for example, the contributions of the gluons is 
$F_S = \del_\mu A_\nu - \del_\nu A_\mu + g_S[A_\mu,A_\nu]$. 
The correct normalization for the interacting terms follows from gauge invariance, and
can be verified directly, e.g. for the gluons, where $\frac 1{4g_{\rm YM}^2} \tr_N (\cF \cF)_G$
gives rise to
\begin{align}
 \frac 1{4g_{\rm YM}^2} \tr_N (g_{\rm YM}^4 [A\tilde t,A\tilde t][A\tilde t,A\tilde t])_G   
 &=  \frac 1{4} \tr_{\rm red} (g_{S}^2 [At, At][A t,A t])_G  \ .
\end{align}
 Finally consider the action for the scalar fields $\Phi^a$, which describe the internal branes
 and contain in particular the Higgs. 
Their kinetic term
 \begin{align}
-\int d^4 x\, \frac 12 \tr_N (D_\mu\Phi^a D^\mu \Phi^a) 
   \label{kinetic-scalars}
 \end{align}
 leads as usual to SSB of the gauge fields.
 The most interesting part is the electroweak symmetry breaking, induced by the 
 minimal fuzzy ellipsoids $S^2_N$. Let us elaborate their effect on the low-energy fields.
These terms arise from 
\begin{align}
S_\phi[A] &= -\frac 12 \int d^4 x\, G^{\mu\nu}\tr_N\Big(
  D_\mu \Phi^4 D_\nu \Phi^4 + D_\mu \Phi^5 D_\nu \Phi^5 \Big) \nn\\
 &=  -\frac 12 \int d^4 x\, G^{\mu\nu}\tr_N\big((D_\mu \Phi^+)^\dagger D_\nu \Phi^+\big)
\end{align}
In view of \eq{EW-branes}, it is natural to organize the non-vanishing entries of $X^+ = X^4 + i X^5$ in terms of 
``effective'' Higgs doublets
\begin{align}
  H_d = \ \begin{pmatrix}
            0 \\ \phi_d
          \end{pmatrix}, \qquad 
  H_u = \  \begin{pmatrix}
          \phi_u \\ 0
          \end{pmatrix}
\end{align}
with  $Y(H_d) = 1$ as in the standard model, 
and  $Y(H_u) = - 1$  as in the MSSM.
Their eigenvalues under $[t_5,.]$ are $+2$.
The scalar fields $\phi$ have dimensions $L^{-1}$ in this section, absorbing 
the scale factor $\frac{\L_0^2}{\pi}$ \eq{field-identification} in their definition. 
 Moreover we will set $\phi_u=\phi_d$ for the VEV's due to
the  relation \eqref{eom-solution-2}, 
which implies
 \begin{align}
 \tan\b = \frac{\phi_u}{\phi_d} = 1 \ .
\end{align}
Note that this tree level relation holds  at very high energies, before integrating out any of the $\cN=4$ fields.
Then  $S_\phi[A]$ takes the standard form of a mass term arising from the covariant derivative of a 
2-component Higgs $H_d$ in the standard model, supplemented by the contribution from 
a second 2-component Higgs $H_u$,
\begin{align}
 S_\phi[A] &=  -\frac 12 \int d^4 x\, G^{\mu\nu} \tr_N\Big((D_\mu H_d )^\dagger D_\nu H_d 
   + (D_\mu H_u)^\dagger D_\nu H_u \Big) \ .
\end{align}
Here 
\begin{align}
  D H_d &= [\cA,H_d] = g W_a t_a H_d + \frac 12(g'B + 2g_5 B_5) H_d\nn\\   
   &=  \frac{\phi}2 \begin{pmatrix}
         g(W_1+iW_2) \\ - g W_3 + g'B+2g_5B_5
        \end{pmatrix} 
     =  \frac{\phi}2 \begin{pmatrix}
         g(W_1+iW_2) \\ - g_Z Z + 2g_5 B_5
        \end{pmatrix},  \nn\\
 D H_u &= g W_a t_a H_d +\frac 12(-g' B + 2g_5B_5)H_u \nn\\
   &=  \frac{\phi}2 \begin{pmatrix}
         g(W_1-iW_2) \\ g W_3-g'B+2g_5B_5
        \end{pmatrix} 
    =  \frac{\phi}2 \begin{pmatrix}
         g(W_1-iW_2) \\  g_Z Z+2g_5B_5
        \end{pmatrix} \ .
\end{align}
The $Z$ boson is identified as the combination of $W_3$ and $B$ which acquires a mass, 
\begin{align}
 g_Z Z = g W_3 - g'B .
\end{align}
On the other hand, the last form of \eq{el-charge} guarantees that $t_Q$ does not couple to the Higgs. 
The masses are obtained from\footnote{The contraction of the vector fields with $G^{\mu\nu}$ is understood.}
\begin{align}
S_\phi[A] 
 &=  - \int d^4 x\, \tr_N\Big(\frac{\phi^2}4 g^2 (W_1^2 + W_2^2) 
   + \frac{\phi^2}4 (g^2+{g'}^2) Z^2 + \phi^2 g_5^2 B_5^2\Big) \ .
 \label{ew-boson-mass}
\end{align}
We can then read off the $W$ and $Z$ bosons masses in the high energy regime, taking into account 
a factor $N_1$ from $\tr_N$:
\begin{align}
 m_W^2 &= \frac 12 N_1 g^2 \phi^2 , \qquad
 m_Z^2 = \frac 12 N_1 (g^2+{g'}^2)\phi^2  \nn\\
 m_5^2 &= 2 N_1 g_5^2 \phi^2 .
 \label{W-mass}
\end{align}
The $U(1)_5$ is anomalous at low energies, hence it is expected to 
disappear from the 
low-energy spectrum  by some St\"uckelberg-type mechanism, cf. 
\cite{Preskill:1990fr,Coriano':2005js,Morelli:2009ev}.
The photon and the $Z$-boson are then identified as usual 
\begin{align}
   \begin{pmatrix}
    A\\  Z
   \end{pmatrix} 
   &=  \begin{pmatrix}
                 \sin \theta_W& \cos\theta_W\\
                 \cos\theta_W & -\sin\theta_W
                \end{pmatrix} 
                 \begin{pmatrix}
                 W_3 \\ B
                \end{pmatrix}  
= \frac{1}{\sqrt{g^2+{g'}^2}} \begin{pmatrix}
                 g' & g\\
                 g & -{g'}
                 \end{pmatrix}
                \begin{pmatrix}
                 W_3 \\ B
                \end{pmatrix}  \ .
\end{align}
 This gives the Weinberg angle
 \begin{align}
  \tan \theta_W &= \frac{g'}{g} = \frac {2 c_{L}}{c_Y}
   = \frac{1}{\sqrt{1+\frac{2N_2}{3N_1}}}
\end{align}
and
\begin{align}
 \sin^2\theta_W = \frac{1}{1+\frac{g^2}{{g'}^2}} = \frac{1}{2+\frac{2N_2}{3 N_1}} \ .
\end{align}
For $N_1=N_2$ this gives $\sin^2\theta_W = 3/8$ and $g_S = g$, as in the $SU(5)$ GUT.

Similarly, we can compute the mass of the gauge bosons $C_\mu$
associated with the breaking of $SU(3) \to SU(2)_L \times U(1)$ due to $r>0$. A typical
generator $t_C \sim |+\rangle_d\langle-|_{u,d} \one_{N_1}$ links the standard model 
fermions to the first massive mirror fermions, such as $\tilde d_R \leftrightarrow d_L, u_L$
or $\tilde e_R \leftrightarrow  \nu_L, e_L$.
Since $t_C$ relates different eigenvalues 
of $X_6$, we have $[X_6,t_C] = \pm r\, t_C$. 
There will also be a mass contribution  from the Higgs $\phi$
in the same way as the $W$ bosons, leading to a mass term 
\begin{align}
 S_{X_6}[C] = -\frac 12 \int d^4 x\, \tr_N\big(g^2 (r^2 + \frac{\phi^2}4) C^\mu C_\mu  \big)
 \label{C-boson-mass}
\end{align}
This is larger than the $W_\pm$ mass, assuming $r>\frac\phi 2$. 

Next consider the contribution to the gauge boson masses 
from the singlet Higgs $S$ via $X^a_{(S)} = h(e^8+ie^9) S + h.c.$. 
Recalling \eq{S-ansatz} and \eq{S-polarization}, the relevant terms in the action are
\begin{align}
 S_{S}[A] &= -\frac 12 \int d^4 x\,\tr_N\big(\sum_{a=8}^9 D_\mu\Phi^a_{(S)} D^\mu\Phi^a_{(S)}\big) \nn\\
  &= -2 \int d^4 x\, h^2\tr_N\big( D_\mu S^\dagger D^\mu S\big) 
 \label{S-mass-contrib}
\end{align}
dropping possible fluctuations of $S$ here. This gives a mass
to every gauge field coupling to $S$. It breaks
 the lepton number $U(1)_l$, and in the absence of $r,\phi$ it breaks
the electroweak $U(4)$ to $U(3)$ (subsequently broken to $SU(2)_L$ by \eq{C-boson-mass}).
In particular $S$ breaks $U(1)_{B-l}$, which is anomaly free and would otherwise 
lead to an unphysical massless gauge boson, whereas $U(1)_l$ is anomalous 
and expected to disappear from the low-energy spectrum anyway.

Finally, the fermion masses for the off-diagonal fermions arise from 
\begin{align}
 \int d^4 x\, g_{\rm YM} \tr_N\bar\psi\Gamma^a [\Phi_a,.]\psi 
  &= 2\int d^4 x\, g_{\rm YM}\tr_N\bar\psi_{12}\Gamma^a [\Phi_a,.]\psi_{12}  \nn\\
   &= 2\int d^4 x\, g_{\rm YM} f_\psi \phi\bar\psi_{12} \psi_{12} \ ,
\end{align}
taking into account the factor 2 from \eq{fermionic-action-block}, which also enters the kinetic term.
Here $f_\psi$ is the Yukawa coupling for the fermion under consideration, 
as discussed in section \ref{sec:Yukawas}.
The trace $\tr_N$ gives no extra factor since the fermions are made from coherent states. Therefore
the fermion mass is given by
\begin{align}
 m_\psi \sim  g_{YM}\phi f_\psi 
\end{align}
where $f_\psi$ is the corresponding Yukawa coupling. For the first series of  mirror fermions
we found $\widetilde f_\psi \approx 1$, so that their tree level (!) mass is about $\sqrt{2}$ times the 
$W$ mass. In contrast, the standard model fermions have much smaller Yukawas. 

At first sight, the low scale of the mirror fermions seems very bad. However, keep in mind that 
we merely computed the tree level masses here\footnote{The term $V_{\rm quant}$ \eq{eq:S_mod} 
in the effective potential does not affect the mass of the fermions and gauge bosons.},  valid at high energies
in the $\cN=4$ regime. 
At lower energies, the Yukawa couplings will be subject to quantum corrections.
For example, an effective factor $\a>1$ in front of the internal Dirac operator $\a\slashed D_{(\rm int)}$ 
rises the fermion masses without affecting the  boson masses, thus
increasing the gap between the electroweak scale and the first mirror fermions. 
More specifically, since the fermions are given by localized (coherent) states on the large branes $S^2_N$,
they couple to all the massive KK gauge and scalar fields arising on these branes.
These KK modes start at a scale set by $R$, which is comparable to the scale of the first mirror fermions $\widetilde \psi$
by \eq{eom-solution-2}.
Therefore they will contribute significantly to the Yukawa couplings.
In contrast, these KK modes do not contribute to the mass of the $W,Z$ gauge bosons, because 
these are $\propto \one$ on the large branes. 
This should magnify the gap between the electroweak scale and the first mirror fermions,
and one may hope that the model can become phenomenologically viable in this way.

In any case, the model clearly predicts mirror fermions with opposite chirality at 
not very high energies. These mirror fermions interact with the standard model gauge bosons, and 
can decay into the standard model fermions
via the heavy gauge bosons $C_\mu$. More quantitative statements would require computing
quantum effects.

\subsection{Moduli and the Higgs potential}
\label{sec:moduli-and-Higgs}

The action for the geometrical moduli $\phi, r, R_i$ is obtained from the modified matrix model action \eqref{eq:S_mod} as
\begin{align}
 S[r,\phi,R] &= -\int d^4 x\, (V_{\rm quant}(r,\phi,R) + V_{\rm int}(r,\phi,R)),   \nn\\
  V_{\rm quant}(r,\phi,R) &= \rho f\left(\pi \rho^{\frac{1}{2}} g_{\YM}^{-1} ( ({R'_u}^2+2R_u^2) 2 c_{N_u} + (r_u^2+2 \phi_u^2) N_u c_{2} + ({R'_l}^2+2R_l^2) c_{N_l})\right), \nn \\
  V_{\rm int}(r,\phi,R) & = \frac{\rho^2}{4 g_{\YM}^2} 2 \left( (2 {R'_u}^2 R_u^2 + R_u^4) 2 c_{N_u} + (2 \phi_u^2 r_u^2 + \phi_u^4) N_u c_2 +  (2 {R'_l}^2 R_l^2 + R_l^4) c_{N_l}\right), \nn
\end{align}
where
\begin{align}
 c_N = \tr_N L_3^2 = \sum_{m=-(N-1)/2}^{(N-1)/2} m^2 = \frac 1{12}N(N^2-1)  \ . \nn
\end{align}
We have
\begin{align*}
 V(r,\phi + \delta \phi,R) & = V(r,\phi,R) + \left( 2 \pi \rho^{\frac{3}{2}} g_{\YM}^{-1} \phi f' +\rho^2 g_{\YM}^{-2} (\phi r^2 + \phi^3) \right) N_u \delta \phi \\
 & + \tfrac{1}{2} \left( 2 \pi \rho^{\frac{3}{2}} g_\YM^{-1} f' + 4 \pi^2 \rho^2 g_\YM^{-2} \phi^2 f'' + \rho^2 g_\YM^{-2} (r^2 + 3 \phi^2) \right) N_u \delta \phi^2 + \cO(\delta \phi^3).
\end{align*}
The coefficient of the term of order $\delta \phi$ vanishes, by the equation of motion \eqref{eom-solution-2}. The remainder can be simplified, and comparison with the kinetic term yields the following mass squared for the fluctuations $\delta \phi$ (here we introduced physical units):
\[
 m^2 = 2 g_\YM^{2} \phi^2 \left( 1 + 2 \pi^2 f'' \right).
\]
This is positive and somewhat larger than $m_W^2$, unless $f''$ is too negative.

There is another interesting set of low-energy perturbations, given by Goldstone bosons
of the global $SO(6)$ symmetry acting uniformly on all matrices, 
corresponding to local rotations of the matrix background.
This affects only the trace-$U(1)$ sector of the model and leads to metric perturbations
related to the effective or ''emergent`` gravity on $\R^4_\theta$,
as elaborated in  \cite{Polychronakos:2013fma} for a similar type of background.

\subsection{Further aspects}

\paragraph{Anomalies and massive gauge fields.}

In the present type of background
(as in analogous brane-configurations in string theory \cite{Antoniadis:2000ena}), 
 a $U(1)$ gauge symmetry arises on each brane, some of which are anomalous at low energies.
 This does not signal an inconsistency, since the fundamental $U(N)$
gauge symmetry is anomaly-free. Rather, it  indicates 
that the corresponding anomalous gauge bosons acquire a mass and disappear from the low-energy physics.
This topic has been discussed extensively in the literature, 
see e.g. \cite{Preskill:1990fr}, 
or \cite{Coriano':2005js,Morelli:2009ev} in a closely related context, based on 
a type of St\"uckelberg mechanism with an axion.

In fact,  axion-like fields 
appear in non-commutative gauge theory via the term $\int \eta(x) F\wedge F$, where the ''axion`` is realized by
the geometric field $\eta(x) = G g$ in the picture of emergent gravity
\cite{Steinacker:2010rh,Steinacker:2008ya,Blaschke:2010qj,Steinacker:2007dq}.
This should be related to the Chern-Simons-terms arising in the 
D-brane action in string theory.
The precise origin of such mass terms in the present context should be clarified.

In particular, baryon number $U(1)_B$ is such an anomalous gauge symmetry. It should still provide protection from
proton decay, in contrast to  many grand-unified models.
This is important in view of the highly populated spectrum of fields at 
intermediate energies.

\paragraph{Generations.}

 Additional generations can arise if the large fuzzy spheres $S^2_{N}$ in either $\cD_{u,d}$
 or $D_{l,c}$ are replaced by stacks of spheres with slightly different parameters. 
On the other hand, we have seen that there are in fact two separated intersection regions
contributing to e.g. $\cD_{u} \cap D_{l}$.  This would also manifest itself as doubling of generations,
which is actually unwelcome as it would imply an even number of generations. 
However,  one of these intersection regions could be removed in principle.

\paragraph{Right-handed neutrinos.}

One clear prediction of our solution is the presence of right-handed neutrinos $\nu_R$,
which acquire a Dirac mass term determined by the corresponding Yukawa coupling. In addition, it seems
plausible that a Majorana mass term \eq{majorana-mass-nu} is induced by quantum effects.
This aspect should  be studied in more detail. 
For a survey on the phenomenological aspects of right-handed neutrinos we refer to the recent review 
\cite{Drewes:2013gca}.


\section{Discussion and conclusion}

We have shown that the IKKT model can behave very similar to the standard model
at low  energies, for suitable backgrounds. 
We provided such backgrounds consisting of branes in the internal space, which are solutions 
of the  matrix model, 
assuming a suitable stabilizing term in the effective potential and a non-linear stabilization of the singlet Higgs.
Our results also apply to  $\cN=4$ $SU(N)$ SYM with sufficiently large $N$, 
challenging the standard lore that $\cN=4$ SYM can only be a ''spherical cow`` approximation
to realistic gauge theories. 
We recover the chiral fermions of the standard model with the correct quantum numbers coupling appropriately to the 
electroweak gauge fields. 
Right-handed neutrinos arise, as well as towers of massive Kaluza-Klein modes of other fields,
ultimately completing the full $\cN=4$ spectrum at very high energies. 
Our results are supported by numerical computations of the spectrum of
low eigenvalues of the internal Dirac operator $\slashed{D}_{\rm int}$, verifying also the chirality and
localization properties of the corresponding fermionic modes.

One clear prediction  is the existence of mirror fermions at intermediate energies, which can decay
to standard model fermions via massive gauge bosons. The mass of the lowest mirror fermions is
rather low at tree level (about $\sqrt{2}$ times the $W$ mass, which  obviously would  not be realistic), however
it seems likely that quantum effects raise their scale to higher energies; this should be studied
in detail elsewhere. They become massless as the Higgs is switched off, reflecting the non-chiral 
nature of the underlying $\cN=4$ theory.

The Higgs sector is found to be more intricate than in the standard model,
consisting of two doublets, which form an intrinsic part
of the internal  branes. This should lead to some protection from quantum corrections. 
The electroweak scale is set by the geometrical scale of the internal compact branes. 
Another important parameters is the rank $N$ of the internal matrices, which determines
the size of the ''large`` internal fuzzy spheres, and  in particular the  
hierarchy of the Yukawa couplings for the standard model fermions versus the mirror fermions. 
The standard model Yukawas can be made arbitrarily small for $N\to\infty$, while those of the mirror sector
remain fixed at tree level. 
However, to assess the viability of the resulting model, quantum corrections 
due to the towers of massive modes must be taken into  account.

There are many open questions and issues raised by this work. One important issue is  the 
scale of the  mirror fermions. 
A reliable computation of this and other physical parameters requires computing the 
quantum corrections due to integrating out the Kaluza-Klein tower of massive fields.
Due to the rich spectrum this is a formidable task even at one loop, 
which however should  be feasible. 
Understanding the quantum contributions to the effective potential is also essential 
to clarify the stability of the background, and to clarify whether it is necessary 
to add a stabilizing term such as \eq{eq:S_mod} by hand.
Ultimately, this should also allow to select preferred backgrounds 
among the mini-landscape of matrix model configurations.

Although we have intentionally hidden any noncommutative aspects, it should be clear that 
our solution really defines a fully noncommutative version of the (extended) standard model
on quantized Minkowski space $\R^4_\theta$. 
If the required stabilizing potential indeed arises through quantum effects in $\cN=4$ SYM, 
the model can be expected to be perturbatively 
finite and free of pathological UV/IR mixing, in contrast to previous proposals 
\cite{Calmet:2001na}.
Moreover, gravity is expected to be included automatically in the matrix model, 
encoded in the trace-$U(1)$ sector \cite{Steinacker:2007dq,Steinacker:2010rh}, 
\cite{Yang:2010kj}. However, this is not yet fully understood. 

Another open issue is the assumed non-linear stablilization of the singlet Higgs $S$. A related aspect is the possible Majorana mass term for $\nu_R$, which is expected to arise due to $S$.


If the above issues can be resolved in a satisfactory way, 
many interesting physical issues could be addressed, including in particular the physical properties 
of the Higgs. 
In any case, we have certainly demonstrated that there is no fundamental obstacle for obtaining
near-standard-model physics from the matrix model.

\paragraph{Acknowledgments.}

This work is supported by the Austrian Science  Fund (FWF): P24713.
Useful discussions with J. Nishimura and A. Tsuchyia are gratefully acknowledged. 

\appendix

\section{Clifford algebra and reduction to 4 dimensions}
\label{sec:clifford}

The ten-dimensional Clifford algebra, generated by
$\Gamma_A$, naturally separates into
a four-dimensional and a six-dimensional one as follows,
\bea
\Gamma_A&=&(\Gamma_{\mu},\Gamma_{3+a}), \nn\\ \Gamma_{\mu}&=&\gamma_
{\mu}\otimes\one_8, \qquad \Gamma_{3+a} = \gamma_5\otimes\Delta_a.
\eea
Here the $\gamma_{\mu}$ define the four-dimensional Clifford algebra, and are chosen 
to be real 
corresponding to the Majorana representation in four dimensions
for $\eta_{\mu\nu} = \diag(-1,1,...,1)$. 
Then  $\gamma_0 = -\gamma_0^\dagger= - \gamma_0^T$ and
$\gamma_i = \gamma_i^\dagger=  \gamma_i^T$.
The $\Delta_a$ define the six-dimensional Euclidean Clifford
algebra, and are chosen to be real and antisymmetric.
The ten-dimensional chirality operator 
\be
\Gamma = \gamma_5 \otimes\Gamma^{(\rm int)}
\ee
separates into four- and six-dimensional chirality operators 
\bea
\gamma_5 &=& -i\gamma_0 ... \gamma_3 = \gamma_5^\dagger
= - \gamma_5^T, \nn\\
\Gamma^{(\rm int)} &=& -i\Delta_1 ... \Delta_6 = (\Gamma^{(\rm int)})^\dagger
= -(\Gamma^{(\rm int)})^T.
\label{Gamma-int}
\eea
Let us denote the ten-dimensional charge conjugation operator as
\be
\cC = C^{(4)} \otimes C^{(6)}, 
\ee
where $C^{(4)}$ is the four-dimensional charge
conjugation operator and $C^{(6)}=\one_8$ in our conventions. This operator 
satisfies, as usual, the relation
\be 
\cC\Gamma^M\cC^{-1}=-(\Gamma^M)^T. 
\ee 
Then
the Majorana condition in 9+1 dimensions is\footnote{Note that $T$ transposes only the spinor.}
$\Psi^C = \cC \obar\Psi^T = \Psi$,
 hence
\begin{align}
\Psi^\star &= \Psi , \qquad 
 \obar\Psi = \Psi^T \cC 
\end{align}
since $\cC = C^{(4)} = \gamma_0$ in the Majorana representation with real $\gamma_\mu$.
Thus the spinor entries are Hermitian matrices  in a MW basis.
The fermionic action can then be written as
\begin{align}
 \Tr \obar\Psi \Gamma_a[X^a,\Psi] &= \Tr \Psi^T \g_0\big(\slashed{D}_4 + \gamma_5\slashed{D}_{(\rm int)}\big)\Psi
\end{align}
where
\begin{align}
\slashed D_{(\rm int)} = \sum_{a=4}^9\Delta^a [X_a,.], 
\qquad 
\{\slashed D_{(\rm int)},\Gamma^{(\rm int)}\} =0
\label{D-6-chirality}
\end{align}
denotes the Dirac operator on the internal space.
The most general 32-component Dirac spinors  $\Psi$ satisfying the Weyl constraint $ \Gamma \Psi = \Psi$
as well as the Majorana condition $\Psi^* = \Psi$ can then be written as 
\bea
\Psi &=& \sum_{i=1}^4\,\Big(\chi_{L,i} \otimes  \eta_{L,i} + (\chi_{L,i} \otimes  \eta_{L,i})^* \Big) , 
\label{Psi-4D}
\eea
where $\chi_{L,i}$ are left-handed  spinors of $\mso(3,1)$, labeled by 
4 left-handed Weyl spinors  $\eta_{L,i}$  of $\mso(6)\cong \msu(4)$.
In particular for spinors taking value in some block-matrix as above, the Majorana condition 
amounts to
\begin{align}
  \Psi   &= \begin{pmatrix}
         \psi_{11} &  \psi_{12} \\
          \psi_{12}^*  &  \psi_{22} 
        \end{pmatrix} , \qquad  \psi_{ii}^* = \psi_{ii}  .
\end{align}
This means that the lower-diagonal matrices $\psi_{12}^*$ are
nothing but ``anti-particles'' of the upper-diagonal ``particles'', so that there is no 
further doubling of the chiral fermions $\psi_{12}$ stretching from brane $1$ to brane $2$ as 
identified in the text.
Spelling out this block-matrix structure of the fermions, the 
Yukawa couplings are 
\begin{align}
 \Tr \obar\Psi\g_5\slashed{D}_{\rm int} \Psi
 &= \Tr\Big( \psi_{11}^T \g_0\g_5 \Delta_c[X^c_{(1)},\psi_{11}]
  + \psi_{22}^T \g_0\g_5 \Delta_c[X^c_{(2)},\psi_{22}]  \nn\\
 &\quad  +  \psi_{12}^T \g_0\g_5 \Delta_c(X^c_{(2)} \psi_{12}^* - \psi_{12}^* X^c_{(1)})
   +  \psi_{12}^{*T}\g_0\g_5 \Delta_c(X^c_{(1)}\psi_{12} -\psi_{12} X^c_{(2)}) \Big) 
 \end{align}
 where $X^c = \diag(X^c_{(1)}, X^c_{(2)})$.
Two of these terms coincide
\begin{align}
  \Tr \psi_{12}^{*T}\g_0\g_5 X^c_{(1)}\Delta_c\psi_{12} 
  = -\Tr X^c_{(1)} \psi_{12}^T \Delta_c\g_0\g_5  \psi_{12}^{*}
\end{align}
using the Grassmann nature of $\psi$, so that
\begin{align}
 \Tr \obar\Psi\g_5\slashed{D}_{\rm int} \Psi
 &= \Tr\Big( \psi_{11}^T \g_0\g_5 \slashed{D}_6 \psi_{11}
  + \psi_{22}^T \g_0\g_5 \slashed{D}_{\rm int} \psi_{22}  
   +  2\psi_{12}^{*T} \g_0\g_5 \slashed{D}_{\rm int} \psi_{12} \Big)
  \label{fermionic-action-block}
\end{align}
This gives rise to the Yukawa couplings computed in section \ref{sec:Yukawas}.


\begin{thebibliography}{99}



\bibitem{Ishibashi:1996xs}
  N.~Ishibashi, H.~Kawai, Y.~Kitazawa, A.~Tsuchiya,
  ``A Large N reduced model as superstring,''
  Nucl.\ Phys.\  {\bf B498 } (1997)  467-491.
  [hep-th/9612115].

  \bibitem{Banks:1996vh} 
  T.~Banks, W.~Fischler, S.~H.~Shenker and L.~Susskind,
  ``M theory as a matrix model: A Conjecture,''
  Phys.\ Rev.\ D {\bf 55}, 5112 (1997)
  [hep-th/9610043];
  B.~de Wit, J.~Hoppe and H.~Nicolai,
  ``On the Quantum Mechanics of Supermembranes,''
  Nucl.\ Phys.\ B {\bf 305}, 545 (1988).

\bibitem{Aoki:1999vr}
  H.~Aoki, N.~Ishibashi, S.~Iso, H.~Kawai, Y.~Kitazawa and T.~Tada,
  ``Noncommutative Yang-Mills in IIB matrix model,''
  Nucl.\ Phys.\ B {\bf 565} (2000) 176
  [hep-th/9908141];

 \bibitem{Aoki:1998vn} 
  H.~Aoki, S.~Iso, H.~Kawai, Y.~Kitazawa and T.~Tada,
  ``Space-time structures from IIB matrix model,''
  Prog.\ Theor.\ Phys.\  {\bf 99}, 713 (1998)
  [hep-th/9802085].

\bibitem{Taylor:2001vb}
  W.~Taylor,
  ``M(atrix) theory: Matrix quantum mechanics as a fundamental theory,''
  Rev.\ Mod.\ Phys.\  {\bf 73} (2001) 419
  [arXiv:hep-th/0101126];
  D.~Kabat and W.~I.~Taylor,
  ``Linearized supergravity from matrix theory,''
  Phys.\ Lett.\  B {\bf 426} (1998) 297
  [arXiv:hep-th/9712185];
  
\bibitem{Steinacker:2010rh}
  H.~Steinacker,
  ``Emergent Geometry and Gravity from Matrix Models: an Introduction,''
  Class.\ Quant.\ Grav.\  {\bf 27 } (2010)  133001.
  [arXiv:1003.4134 [hep-th]]
  
\bibitem{Steinacker:2007dq} 
  H.~Steinacker,
  ``Emergent Gravity from Noncommutative Gauge Theory,''
  JHEP {\bf 0712}, 049 (2007)
  [arXiv:0708.2426 [hep-th]].
  

  
\bibitem{Kim:2011cr}
  S.~-W.~Kim, J.~Nishimura, and A.~Tsuchiya,
  ``Expanding (3+1)-dimensional universe from a Lorentzian matrix model for superstring theory in (9+1)-dimensions,''
  Phys.\ Rev.\ Lett.\  {\bf 108} (2012) 011601
  [arXiv:1108.1540 [hep-th]];
  Y.~Ito, S.~-W.~Kim, Y.~Koizuka, J.~Nishimura and A.~Tsuchiya,
  ``A renormalization group method for studying the early universe in the Lorentzian IIB matrix model,''
  arXiv:1312.5415 [hep-th].
  
\bibitem{Chatzistavrakidis:2011gs}
  A.~Chatzistavrakidis, H.~Steinacker and G.~Zoupanos,
  ``Intersecting branes and a standard model realization in matrix models,''
  JHEP {\bf 1109} (2011) 115
  [arXiv:1107.0265 [hep-th]].
  
 \bibitem{Connes:1990qp} 
  A.~Connes and J.~Lott,
  ``Particle Models and Noncommutative Geometry (Expanded Version),''
  Nucl.\ Phys.\ Proc.\ Suppl.\  {\bf 18B}, 29 (1991).
  
  
  \bibitem{Chepelev:1997av} 
  I.~Chepelev and A.~A.~Tseytlin,
  ``Interactions of type IIB D-branes from D instanton matrix model,''
  Nucl.\ Phys.\ B {\bf 511}, 629 (1998)
  [hep-th/9705120].
  
  
  \bibitem{Alishahiha:1999ci} 
  M.~Alishahiha, Y.~Oz and M.~M.~Sheikh-Jabbari,
  ``Supergravity and large N noncommutative field theories,''
  JHEP {\bf 9911}, 007 (1999)
  [hep-th/9909215].
  
  
  
  
  \bibitem{Kitazawa:2005ih}
  Y.~Kitazawa and S.~Nagaoka,
  ``Graviton propagators on fuzzy G/H,''
  JHEP {\bf 0602} (2006) 001
  [hep-th/0512204].
  
  
\bibitem{Blaschke:2011qu} 
  D.~N.~Blaschke and H.~Steinacker,
  ``On the 1-loop effective action for the IKKT model and non-commutative branes,''
  JHEP {\bf 1110}, 120 (2011)
  [arXiv:1109.3097 [hep-th]].
  
\bibitem{Nishimura:2013moa} 
  J.~Nishimura and A.~Tsuchiya,
  ``Realizing chiral fermions in the type IIB matrix model at finite N,''
  JHEP {\bf 1312}, 002 (2013)
  [arXiv:1305.5547 [hep-th]].

  
\bibitem{Steinacker:2013eya} 
  H.~Steinacker and J.~Zahn,
  ``An Index for Intersecting Branes in Matrix Models,''
  SIGMA {\bf 9},  067, 7 pages (2013)
  [arXiv:1309.0650 [hep-th]].
  
\bibitem{Aoki:2010gv}
  H.~Aoki,
  ``Chiral fermions and the standard model from the matrix model compactified on a torus,''
  Prog.\ Theor.\ Phys.\  {\bf 125} (2011) 521
  [arXiv:1011.1015 [hep-th]].
 
  
\bibitem{Berkooz:1996km} 
  M.~Berkooz, M.~R.~Douglas and R.~G.~Leigh,
  ``Branes intersecting at angles,''
  Nucl.\ Phys.\ B {\bf 480}, 265 (1996)
  [hep-th/9606139].
  
\bibitem{Antoniadis:2000ena}
  I.~Antoniadis, E.~Kiritsis, T.~N.~Tomaras,
  ``A D-brane alternative to unification,''
  Phys.\ Lett.\  {\bf B486 } (2000)  186-193.
  [hep-ph/0004214];
 G.~Aldazabal, S.~Franco, L.~E.~Ibanez, R.~Rabadan and A.~M.~Uranga,
  ``Intersecting brane worlds,''
  JHEP {\bf 0102} (2001) 047
  [hep-ph/0011132].
  I.~Antoniadis, E.~Kiritsis, J.~Rizos, T.~N.~Tomaras,
  ``D-branes and the standard model,''
  Nucl.\ Phys.\  {\bf B660 } (2003)  81-115.
  [hep-th/0210263];
  L.~E.~Ibanez, F.~Marchesano, R.~Rabadan,
  ``Getting just the standard model at intersecting branes,''
  JHEP {\bf 0111 } (2001)  002.
  [hep-th/0105155]; 
  C.~Kokorelis,
  ``Exact standard model structures from intersecting D5-branes,''
  Nucl.\ Phys.\ B {\bf 677}, 115 (2004)
  [hep-th/0207234].
  R.~Blumenhagen, M.~Cvetic, P.~Langacker, G.~Shiu,
  ``Toward realistic intersecting D-brane models,''
  Ann.\ Rev.\ Nucl.\ Part.\ Sci.\  {\bf 55 } (2005)  71-139.
  [hep-th/0502005]; 
  F.~Marchesano,
  ``Progress in D-brane model building,''
  Fortsch.\ Phys.\  {\bf 55 } (2007)  491-518.
  [hep-th/0702094].
   
  
 \bibitem{Hoppe:2012}
  J.~Arnlind, J. Choe, J. Hoppe
  ``Noncommutative minimal surfaces''.
  [arXiv:13010757 [math.QA]]
  
  
  \bibitem{Madore:1991bw}
  J.~Madore,
  ``The fuzzy sphere,''
  Class.\ Quant.\ Grav.\  {\bf 9} (1992) 69.
 
  
\bibitem{hoppe}
 J. Hoppe, "Quantum theory of a massless relativistic surface and a two-dimensional bound state problem``, 
 PH D thesis, MIT 1982;
J.~Hoppe,``Membranes and matrix models,''
  [hep-th/0206192].
  
  
\bibitem{Grosse:2010zq}
  H.~Grosse, F.~Lizzi, H.~Steinacker,
  ``Noncommutative gauge theory and symmetry breaking in matrix models,''
  Phys.\ Rev.\  {\bf D81 } (2010)  085034.
  [arXiv:1001.2703 [hep-th]].
  
  
  \bibitem{Aschieri:2006uw} 
  P.~Aschieri, T.~Grammatikopoulos, H.~Steinacker and G.~Zoupanos,
  ``Dynamical generation of fuzzy extra dimensions, dimensional reduction and symmetry breaking,''
  JHEP {\bf 0609}, 026 (2006)
  [hep-th/0606021].
  
  
\bibitem{Morelli:2009ev} 
  S.~Morelli,
  ``St\"uckelberg Axions and Anomalous Abelian Extensions of the Standard Model'',
  PhD thesis Salento,
  arXiv:0907.3877 [hep-ph].

  \bibitem{Coriano':2005js} 
  C.~Coriano, N.~Irges and E.~Kiritsis,
  ``On the effective theory of low scale orientifold string vacua,''
  Nucl.\ Phys.\ B {\bf 746}, 77 (2006)
  [hep-ph/0510332].

\bibitem{Preskill:1990fr} 
  J.~Preskill,
  ``Gauge anomalies in an effective field theory,''
  Annals Phys.\  {\bf 210}, 323 (1991).
  
 
\bibitem{Steinacker:2011wb}
  H.~Steinacker,
  ``Split noncommutativity and compactified brane solutions in matrix models,''
  Prog.\ Theor.\ Phys.\  {\bf 126} (2012) 613
  [arXiv:1106.6153 [hep-th]].

  
\bibitem{Hoppe:1997gr} 
  J.~Hoppe, 
  ``Some classical solutions of membrane matrix model equations,''
  In *Cargese 1997, Strings, branes and dualities* 423-427
  [hep-th/9702169];

\bibitem{Polychronakos:2013fma} 
  A.~P.~Polychronakos, H.~Steinacker and J.~Zahn,
  ``Brane compactifications and 4-dimensional geometry in the IKKT model,''
  Nucl.\ Phys.\ B {\bf 875}, 566 (2013)
  [arXiv:1302.3707 [hep-th]].
  
  
\bibitem{Perelomov:1986tf}
  A.~M.~Perelomov,
  ``Generalized coherent states and their applications,''
  Berlin,  Springer (1986);
   H.~Grosse, P.~Presnajder,
  ``The Construction of noncommutative manifolds using coherent states,''
  Lett.\ Math.\ Phys.\  {\bf 28 } (1993)  239-250.
 
 
\bibitem{Berenstein:2012ts} 
  D.~Berenstein and E.~Dzienkowski,
  ``Matrix embeddings on flat $R^3$ and the geometry of membranes,''
  Phys.\ Rev.\ D {\bf 86}, 086001 (2012)
  [arXiv:1204.2788 [hep-th]].
  
 
  
  
\bibitem{Steinacker:2008ya} 
  H.~Steinacker,
  ``Covariant Field Equations, Gauge Fields and Conservation Laws from Yang-Mills Matrix Models,''
  JHEP {\bf 0902}, 044 (2009)
  [arXiv:0812.3761 [hep-th]].

  \bibitem{Blaschke:2010qj} 
  D.~N.~Blaschke and H.~Steinacker,
  ``Curvature and Gravity Actions for Matrix Models II: The Case of general Poisson structure,''
  Class.\ Quant.\ Grav.\  {\bf 27}, 235019 (2010)
  [arXiv:1007.2729 [hep-th]].
  
  
\bibitem{Drewes:2013gca} 
  M.~Drewes,
  ``The Phenomenology of Right Handed Neutrinos,''
  Int.\ J.\ Mod.\ Phys.\ E {\bf 22}, 1330019 (2013)
  [arXiv:1303.6912 [hep-ph]].
  
  
 \bibitem{Calmet:2001na} 
  X.~Calmet, B.~Jurco, P.~Schupp, J.~Wess and M.~Wohlgenannt,
  ``The Standard model on noncommutative space-time,''
  Eur.\ Phys.\ J.\ C {\bf 23}, 363 (2002)
  [hep-ph/0111115];
  V.~V.~Khoze and J.~Levell,
  ``Noncommutative standard modelling,''
  JHEP {\bf 0409}, 019 (2004)
  [hep-th/0406178];
 M.~Chaichian, P.~Presnajder, M.~M.~Sheikh-Jabbari and A.~Tureanu,
 ``Noncommutative standard model: Model building,''
  Eur.\ Phys.\ J.\ C {\bf 29}, 413 (2003)
  [hep-th/0107055].
  
  

 \bibitem{Yang:2010kj} 
  H.~S.~Yang,
  ``Emergent Geometry and Quantum Gravity,''
  Mod.\ Phys.\ Lett.\ A {\bf 25}, 2381 (2010)
  [arXiv:1007.1795 [hep-th]].
  
  
  
 
 

 
 
 
 
 
  
%
%
  
  
  
\end{thebibliography}
\end{document}